\documentclass[aps,prd,superscriptaddress,nofootinbib,twocolumn]{revtex4-2}

\usepackage{mathtools}
\usepackage{amsfonts}
\usepackage{mathrsfs}

\usepackage{bbm}
\usepackage{slashed}
\usepackage{amsmath}
\usepackage{bm}
\usepackage{tensor}

\usepackage{graphicx}
\usepackage{color}
\usepackage{colortbl}
\usepackage{array}
\usepackage[dvipsnames]{xcolor}

\usepackage{float}
\usepackage{placeins}
\usepackage{booktabs}
\usepackage{subcaption}
\usepackage{makecell}
\usepackage{tabstackengine}

\usepackage{xspace}
\usepackage{hyperref}
\usepackage[nameinlink]{cleveref}
\usepackage{bookmark}

\usepackage{xifthen}
\usepackage{xcolor}
\usepackage{enumitem}
\hypersetup{
	colorlinks,
	linkcolor={red!75!black},
	citecolor={blue!75!black},
	urlcolor={blue!75!black}
}
\usepackage{units}

\usepackage[utf8]{inputenc}

\setkeys{Gin}{width=0.48\textwidth}

\newcommand*{\ie}{i.e.\@\xspace}

\makeatletter
\newsavebox\myboxA
\newsavebox\myboxB
\newlength\mylenA

\newcommand*\xoverline[2][0.75]{%
	\sbox{\myboxA}{$\m@th#2$}%
	\setbox\myboxB\null
	\ht\myboxB=\ht\myboxA%
	\dp\myboxB=\dp\myboxA%
	\wd\myboxB=#1\wd\myboxA
	\sbox\myboxB{$\m@th\overline{\copy\myboxB}$}
	\setlength\mylenA{\the\wd\myboxA}
	\addtolength\mylenA{-\the\wd\myboxB}%
	\ifdim\wd\myboxB<\wd\myboxA%
	\rlap{\hskip 0.5\mylenA\usebox\myboxB}{\usebox\myboxA}%
	\else
	\hskip -0.5\mylenA\rlap{\usebox\myboxA}{\hskip 0.5\mylenA\usebox\myboxB}%
	\fi}
\makeatother

\newcommand{\imag}{\text{i}}

\newcommand{\tinytext}[1]{\text{\tiny{#1}}}

\graphicspath{{./figures/}}

\usepackage{axodraw2}

\usepackage{xifthen}
\usepackage{xcolor}

\newcommand{\gettitle}{Scalar spectral functions from the spectral fRG}

\hypersetup{
	colorlinks,
	linkcolor={red!75!black},
	citecolor={blue!75!black},
	urlcolor={blue!75!black}, 
	pdftitle={\gettitle},
	pdfauthor={Wessely},
	pdfkeywords={analytic continuation}
	{correlations functions}
	{functional renormalisation group}
	{real time} {spectral function} 
	bookmarksopen=true,
	bookmarksopenlevel=2,
	bookmarksnumbered=true
}

\begin{document}

\title{\gettitle}

\author{Jan~Horak}
  \affiliation{Institut f\"ur Theoretische Physik, Universit\"at Heidelberg, Philosophenweg 16, 69120
  Heidelberg, Germany}

\author{Friederike~Ihssen}
\affiliation{Institut f\"ur Theoretische Physik, Universit\"at Heidelberg, Philosophenweg 16, 69120
	Heidelberg, Germany}  

\author{Jan M.~Pawlowski}
  \affiliation{Institut f\"ur Theoretische Physik, Universit\"at Heidelberg, Philosophenweg 16, 69120
    Heidelberg, Germany}
  \affiliation{ExtreMe Matter Institute EMMI, GSI, Planckstr. 1, D-64291 Darmstadt, Germany}

  \author{Jonas~Wessely}
  \affiliation{Institut f\"ur Theoretische Physik, Universit\"at Heidelberg, Philosophenweg 16, 69120
  	Heidelberg, Germany}
  
\author{Nicolas~Wink}
  \affiliation{Institut f\"ur Kernphysik (Theoriezentrum), Technische Universit\"at Darmstadt, 
  	D-64289 Darmstadt, Germany}

\begin{abstract}
	We compute non-perturbative spectral functions in a scalar $\phi^4$-theory in three spacetime dimensions via the spectral functional renormalisation group. This approach allows for the direct, manifestly Lorentz covariant computation of correlation functions in Minkowski spacetime, including a physical on-shell renormalisation. We present numerical results for the spectral functions of the two- and four-point correlation functions for different values of the coupling parameter. These results agree very well with those obtained from another functional real-time approach, the spectral Dyson-Schwinger equation.
\end{abstract}

\maketitle

\section{Introduction}

We set up the spectral functional renormalisation group (fRG) for a scalar $\phi^4$-theory in three spacetime dimensions. The spectral fRG is a non-perturbative functional real-time approach for the direct computation of correlation functions in Minkowski spacetime. It is based on the general functional real-time setup introduced in~\cite{Horak:2020eng, Horak:2021pfr, Horak:2022myj, Horak:2022aza}, first applied to Dyson-Schwinger equations (DSE). The approach is based on the K\"all\'en-Lehmann spectral representation~\cite{Kallen:1952zz,Lehmann:1954xi} for the two-point function, which allows to analytically access the momentum structure of functional diagrammatic expressions. The setup has been extended to the fRG approach by using a masslike Callan-Symanzik (CS) regulator in~\cite{Braun:2022mgx} and has been applied to gravity in~\cite{Fehre:2021eob}. The CS regulator sustains spectral representations alongside with Lorentz invariance, and allows for a spectral renormalisation consistent with all symmetries at hand; for more details see~\cite{Horak:2020eng, Braun:2022mgx}. Moreover, in~\cite{Braun:2022mgx} the concept of flowing renormalisation has been introduced, which allows for an on-shell renormalisation at each renormalisation group scale. For further real-time applications of the fRG in a broad variety of research fields, see e.g., \cite{Gasenzer:2007za, Gasenzer:2010rq, Floerchinger:2011sc, Kamikado:2013sia, Tripolt:2013jra, Pawlowski:2015mia, Kamikado:2016chk, Jung:2016yxl, Pawlowski:2017gxj, Wang:2017vis, Tripolt:2018jre, Tripolt:2018qvi, Corell:2019jxh, Huelsmann:2020xcy, Jung:2021ipc, Tan:2021zid, Heller:2021wan, Fehre:2021eob, Roth:2021nrd, Roth:2023wbp}.

In the present work, we accompany the conceptual progress made in~\cite{Braun:2022mgx} with a non-perturbative application to spectral functions in the three dimensional \mbox{$\phi^4$-theory}. This allows to directly compare our results with those obtained in~\cite{Horak:2020eng} within the spectral DSE approach. Both functional approaches implement different resummation schemes for the correlators of the given theory through infinite towers of one-loop (fRG) or two-loop (DSE) exact diagrammatic relations. Within an fRG implementation, the successive momentum-shell integration of loop momenta $p^2 \approx k^2$ with the infrared cutoff scale $k$, already provides an average momentum dependence within simple approximations. Due to their intricate spectral representation, this is particularly beneficial for including non-trivial vertices into the flow, e.g., via momentum-independent but cutoff-dependent approximations. 

This work is organised as follows: In \Cref{sec:specDiag}, we briefly discuss the spectral functional approach. In \Cref{sec:frg} we set up its application to the functional renormalisation group for a scalar theory. After discussing the different phases of the theory in~\Cref{sec:phi4}, we present our results in \Cref{sec:Results}. This includes a detailed comparison to those obtained with the spectral DSE in \cite{Horak:2020eng}. We summarise our findings in \Cref{sec:Conclusions}.

\section{Spectral functions and functional equations}
\label{sec:specDiag}

\begin{figure*}[t]
	\centering
	\begin{subfigure}{.48\linewidth}
		\centering
		\includegraphics[width=.90\textwidth]{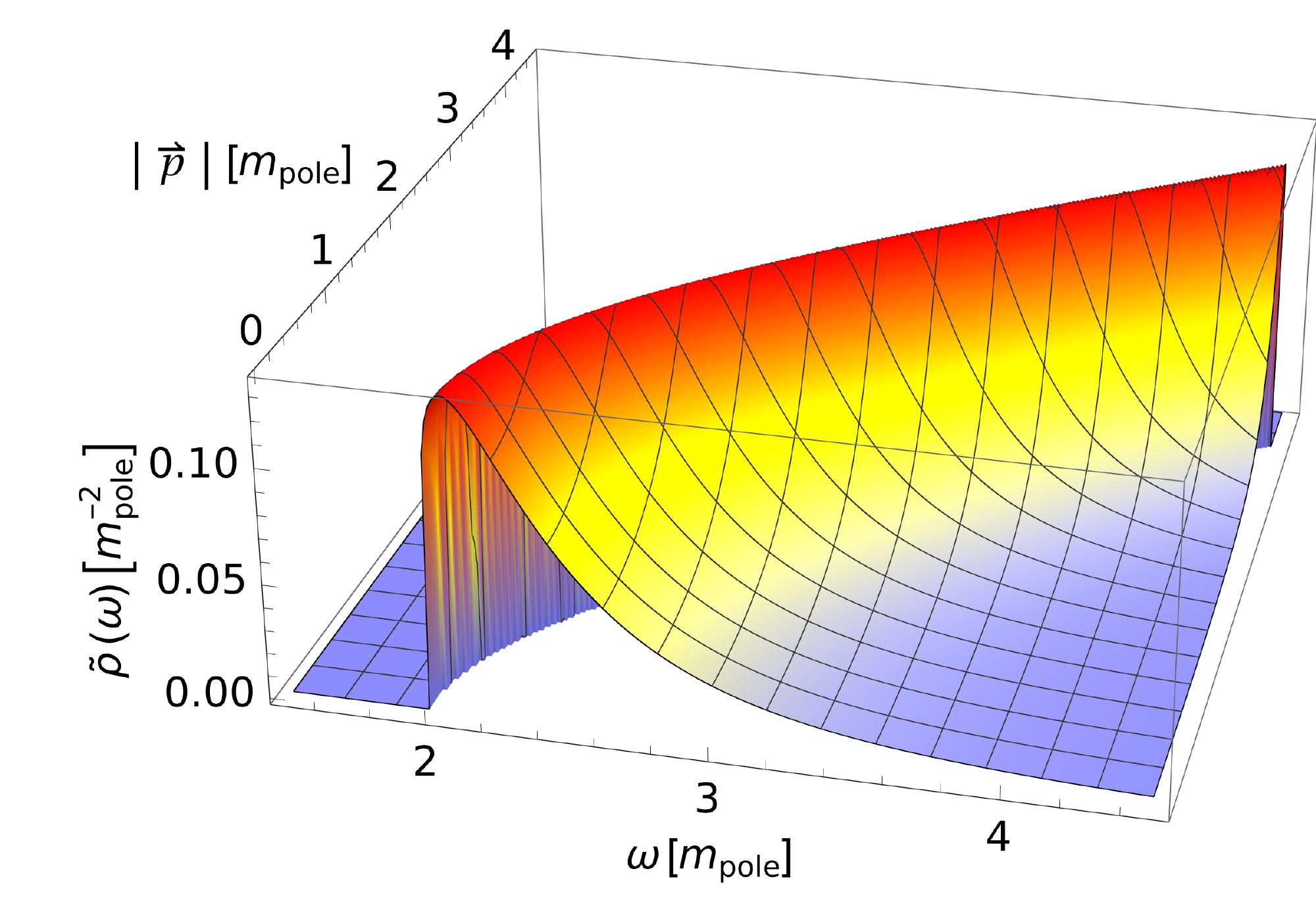}
		\caption{Scattering tail of the propagator spectral function as a function of frequency and spatial momentum in the $(1+2)$-dimensional $\phi^4$-theory in the broken phase. It features a sharp onset at the $1\rightarrow 2$ particle onset and explicit Lorentz invariance. Higher scattering onsets are strongly suppressed.\vspace{3.5mm}\hspace*{\fill}}
		\label{fig:specfuncsSpatial3dprop}
	\end{subfigure}%
	\hspace{0.03\linewidth}%
	\begin{subfigure}{.48\linewidth}
		\centering
			\includegraphics[width=.90\textwidth]{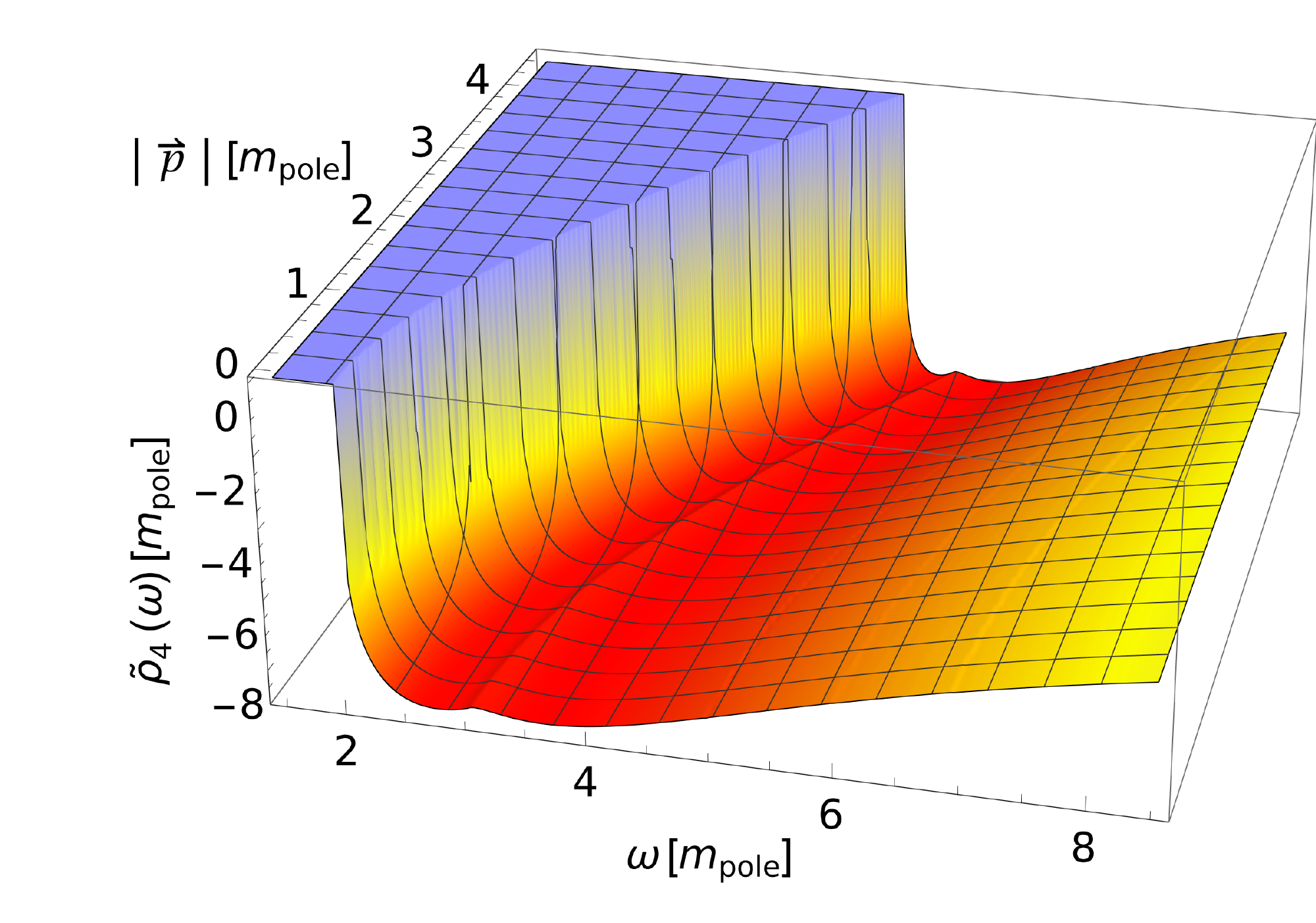}
		\caption{Spectrum of the resummed four-vertex in an $s$-channel approximation as a function of frequency and spatial momentum. It features Lorentz invariance and exhibits a sharp onset at the two-particle threshold. The scattering spectrum also has a visible three-particle onset at $3m_\tinytext{pole}$.\vspace{3.5mm} \hspace*{\fill}}
		\label{fig:specfuncsSpatial3dvertex}
	\end{subfigure}
	\caption{Propagator and vertex scattering spectra in a $(1+2)$-dimensional $\phi^4$-theory in the broken phase. All quantities are measured in units of the pole mass, with a coupling strength $\lambda / m_\tinytext{pole}=20$. \hspace*{\fill}}
	\label{fig:specfuncsSpatial3d}
\end{figure*}

The central idea of spectral functional approaches is to use the spectral representation for all propagators and vertices in the non-perturbative loop diagrams. Then, the momentum integrals can be performed analytically, and the remaining numerical task boils down to the solution of real spectral integrals.

Most of the relations in the present section can be generalised straightforwardly to general field theories. For the sake of simplicity we already restrict ourselves to a $\phi^4$-theory in $(1+2)$ dimensions, for which also the explicit numerical results in the present work are obtained. Its classical action reads  
\begin{align} 
	S[\phi] &=\int\! \mathrm{d}^3x \,\bigg\{ \frac{1}{2} \phi\Bigl(-\partial^2 + \mu \Bigr)\phi + \frac{\lambda_\phi}{4!} \phi^4\bigg\} \,. 
\label{eq:S}
\end{align} 
For $\mu >0$, the minimum of the classical potential is at vanishing field. Then, the mass parameter can be identified with the classical mass squared, $m_\phi^2=\mu$. For $\mu<0$, the full potential exhibits non-trivial minima, and the classical mass of the theory follows from the effective potential as $m_\phi^2=-2 \mu$.

\subsection{Spectral properties of the two-point function}
\label{sec:KL}

The spectral representation of the propagator of a given field $\phi$ is at the core of the spectral functional approach. In the present work, we assume the Källén-Lehmann (KL) representation of the full propagator $G(p_0,\vec p)$ of the field $\phi$ to hold, 
\begin{subequations} \label{eq:specrep}
\begin{align} 
	G(p_0,\vec{p}) = \int_{0_-}^{\infty} \frac{\mathrm{d}\lambda}{\pi} \frac{\lambda \rho\left(\lambda,\vec{p}\,\right)}{\lambda^2 + p_0^2} \,, 
\label{eq:SpecInt}
\end{align}
where $0_-$ ensures that massless poles are taken into account properly. The spectral function $\rho(\lambda)$ is the probability density of creating a Fock state with energy $\lambda$ from the vacuum in the presence of the quantum field $\phi$. It is related to the propagator by 
\begin{align} \label{eq:rho}
	\rho(\omega,\vec{p}\,) = 2 \, \text{Im} \; G \left(p_0 = - \imag (\omega+ \imag 0^+),\vec{p}\,\right) \,.
\end{align} 
\end{subequations}
The propagator is a function of $p^2$ due to Lorentz symmetry. This allows us to drop any explicit $\vec p\,$-dependence from now on and identify $p_0^2 = p^2$.

The spectral function $\rho$ encodes all dynamical, perturbative and non-perturbative information of the propagator. 

In~\labelcref{eq:specrep}, $p_0$ denotes the Euclidean and $\omega$ the Minkowski frequency. In the absence of higher order resonances, the spectral function of the $\phi^4$-theory is given by 
\begin{align}
	\rho(\omega) = \frac{2\pi}{Z_\phi} \delta(\omega^2 - m_\tinytext{pole}^2) + \theta(\omega^2- m_\tinytext{scat}^2)  \tilde \rho(\omega) \,, 
 \label{eq:scale-dep-specIntro}
\end{align}
with $\rho(\omega)=\rho(\omega,0)$, and $\tilde\rho(\omega)=\tilde\rho(\omega,0)$ for the scattering continuum $\tilde\rho$. The mass $m_\tinytext{pole}$ in \labelcref{eq:scale-dep-specIntro} is the pole mass of the full quantum theory, defined by $G^{-1}(\pm m_\tinytext{pole},0)=0$.  

The scattering continuum sets in at $\lambda^2 = m_\tinytext{scat}^2$. In the case of a non-vanishing background field, the theory admits $1\to 2$ scattering (broken phase), and we have $m_\tinytext{scat}=2 m_\tinytext{pole}$. \Cref{fig:specfuncsSpatial3dprop} shows the full scattering tail of the propagator as a function of the frequency $\omega$ and spatial momentum $|\vec p|$ in the broken phase. Higher thresholds of $1\rightarrow N$ scattering processes lead to further discontinuities in the scattering tail and are typically strongly suppressed. In the absence of $1\to 2$ scatterings (symmetric phase), the first allowed scattering is $1\to 3$ and the scattering threshold is $m_\tinytext{scat}=3 m_\tinytext{pole}$, the respective spectral function is depicted in \Cref{fig:PropSpecSym} and discussed there. 

If the spectral representation~\labelcref{eq:specrep} holds, all non-analyticities of the propagator lie on the real frequency axis. These  non-analyticities are given by either poles or cuts. Poles originate from asymptotic states that overlap with the propagator of the field $\phi$, while cuts represent scattering states. 

For the propagator of a physical field that describes an asymptotic state, the spectral function is positive. Furthermore, the canonical commutation relations imply a normalisation via the sum rule,
\begin{align} 
	\int_{0_-}^{\infty} \frac{\mathrm{d}\lambda}{\pi} \,\lambda\, \rho(\lambda,\vec p) = 1 \,, 
\label{eq:specnorm}
\end{align} 
for all spatial momenta. Inserting the spectral function \labelcref{eq:scale-dep-specIntro} into \labelcref{eq:specnorm}, we arrive at
\begin{align}
	\frac{1}{Z_\phi} = 1- \int_{m_\tinytext{scat}}^{\infty} \frac{\mathrm{d}\lambda}{\pi} \,\lambda\, \tilde\rho(\lambda,\vec{p}\,)  \,. 
 \label{eq:specnormForAllk}
\end{align} 
 \Cref{eq:specnormForAllk} comprises the well-known property that the on-shell amplitude $1/Z_\phi$ is bounded from above by unity, $Z_\phi\geq 1$, as the scattering tail carries part of the total probability.

\subsection{Spectral properties of the four-point function}
\label{sec:Spec4point}

Vertices also admit spectral representations, which get increasingly complicated for higher order correlation functions due to their increase in arguments. In the present case, we restrict ourselves to an $s$-channel approximation of the full one-particle irreducible (1PI) four-point function or vertex. This leaves us with a single momentum argument and an accordingly simple spectral representation. The four-point function is given by the fourth field derivative of the effective action $\Gamma[\phi]$, whose $n^{\text{th}}$ field derivatives $\Gamma^{(n)}[\phi]$ are the 1PI $n$-point functions. We use a spectral representation for this $s$-channel vertex~\cite{Horak:2020eng}, 
\begin{align}\nonumber 
	\Gamma^{(4)}(p_0,\vec p\,) =& \, \lambda_{\phi} + \int_\lambda \frac{\rho_{4}(\lambda,\vec p\,)}{\lambda^2 + p_0^2} \,, \\[1ex]
	\rho_{4}(\omega,\vec p\,)=& \, 2 \, \text{Im} \, \Gamma^{(4)}(p_0 = - \imag (\omega + \imag 0^+),\vec p\,) \,,
 \label{eq:specrep4point} 
\end{align}
where $\lambda_\phi$ is the classical vertex in \labelcref{eq:S} and 
\begin{align}
\int_\lambda = \int_{0_-}^{\infty} \frac{\mathrm{d}\lambda}{\pi} \lambda\,.
\end{align}
Analogue to the spectral function of the propagator, $\rho_4$ is defined by the discontinuities of the four-point function, see~\labelcref{eq:specrep4point}. Also for the four-point function, the spatial momentum dependence of spectral function $\rho_{4}(\omega,\vec p\,)$ follows from the one at vanishing spatial momentum, $\rho_{4}(\omega)=\rho_{4}(\omega, 0)$ via a Lorentz boost. 

\Cref{fig:specfuncsSpatial3dvertex} shows the spectrum of the four-point function in the $s$-channel approximation discussed in~\Cref{sec:phi4}. It shows the $2 \rightarrow 2$ scattering onset at twice the pole mass $m_\tinytext{pole}$ of the field $\phi$. The next threshold from the $2 \rightarrow 3$ scattering is also visible, but the result also contains the strongly suppressed threshold of higher order scattering processes. A more detailed discussion of our results is given in~\Cref{sec:Results}.

\subsection{Structural properties of diagrams}
\label{sec:structural_diags}

In the spectral functional approach, spectral representations are utilised to rewrite diagrams in terms of momentum loop integrals over classical propagators with spectral masses and residual spectral integrals; for a general discussion see~\cite{Horak:2020eng}. In the present work, we apply this approach in the context of the functional renormalisation group, amounting to the \textit{spectral fRG} approach detailed in \Cref{sec:frg}. This leads to one-loop exact relations for correlation functions in terms of full propagators and vertices. In addition, we use a one-loop closed, resummed Bethe-Salpeter kernel to compute the four-point function. 

As discussed above, the spectral fRG leads to perturbative one-loop momentum integrals in diagrams, which can be solved analytically. The non-perturbative information of the diagrams such as pole masses and thresholds is stored in the remaining spectral integrals. For the present purpose, it is sufficient to consider a single external momentum argument, which is either that of the propagator or the $s$-channel momentum of the four-point function. However, the generalisation to diagrams with several external momenta as present in the spectral computations of general $n$-point functions is straightforward. 

In the present case, we only have to consider diagrams with one in-flowing or out-flowing external momentum $\pm p$, and we encounter diagrams of the general form  
\begin{align}
	D[p] = g \int_{q} \text{Vert}(p, q) \prod_{j=1}^{N} G(l_j) \, ,
 \label{eq:GenLoop1}
\end{align}
where $l_i=q,q\pm p$ are the momenta of the $N$ propagators and we have used the abbreviation 
\begin{align}
	\int_q = \int \frac{\mathrm{d}^dq}{(2\pi)^d} \,. 
\end{align}
$\text{Vert}(p,q)$ carries the momentum dependence of all vertices, which we assume to be either a polynomial or rational function of $p$ and the $l_i$, or to admit a spectral representation. All prefactors are collected in the overall prefactor $g$. By inserting the spectral representation~\labelcref{eq:specrep} for each propagator, the momentum integrals acquire a standard perturbative form, where the masses are the respective spectral parameters squared, $\lambda_i^2$. Finally, the spectral parameters are integrated over, weighted by the respective spectral function, 
\begin{align}
	D[p]=g\prod_{j=1}^{N} \int_{\lambda_j} \rho(\lambda_j)I(\lambda_1,..,\lambda_N,p) \,,
  \label{eq:spectralDiagGeneral}
\end{align}
with
\begin{align} 
 	I(\lambda_1,..,\lambda_N,p)=\int_{q} \text{Vert}(p,q) \prod_{j=1}^{N} \frac{1}{\lambda_j^2 + l_j^2}\,.
\label{eq:momIntGeneral}
\end{align}
The momentum integral in \labelcref{eq:momIntGeneral} is readily solved and the resulting analytic expression holds true for $p \, \in \, \mathbb{C}$. This gives us access to the spectral function~\labelcref{eq:specrep} via the limit $p\rightarrow -\imag (\omega + \imag 0^+)$. We remark that in the present spectral fRG approach to the \mbox{(1+2)-dimensional} scalar theory, all integrals are finite, and we can safely change the order of integration even prior to renormalisation. In general the interchange of momentum and spectral integration performed in~\labelcref{eq:spectralDiagGeneral} assumes a suitable regularisation of the full integral, which can be done with \textit{spectral renormalisation}~\cite{Horak:2020eng}. 

\begin{figure*}[t]
	\centering
	\begin{subfigure}{.48\linewidth}
		\centering
		\includegraphics[width=.99\textwidth]{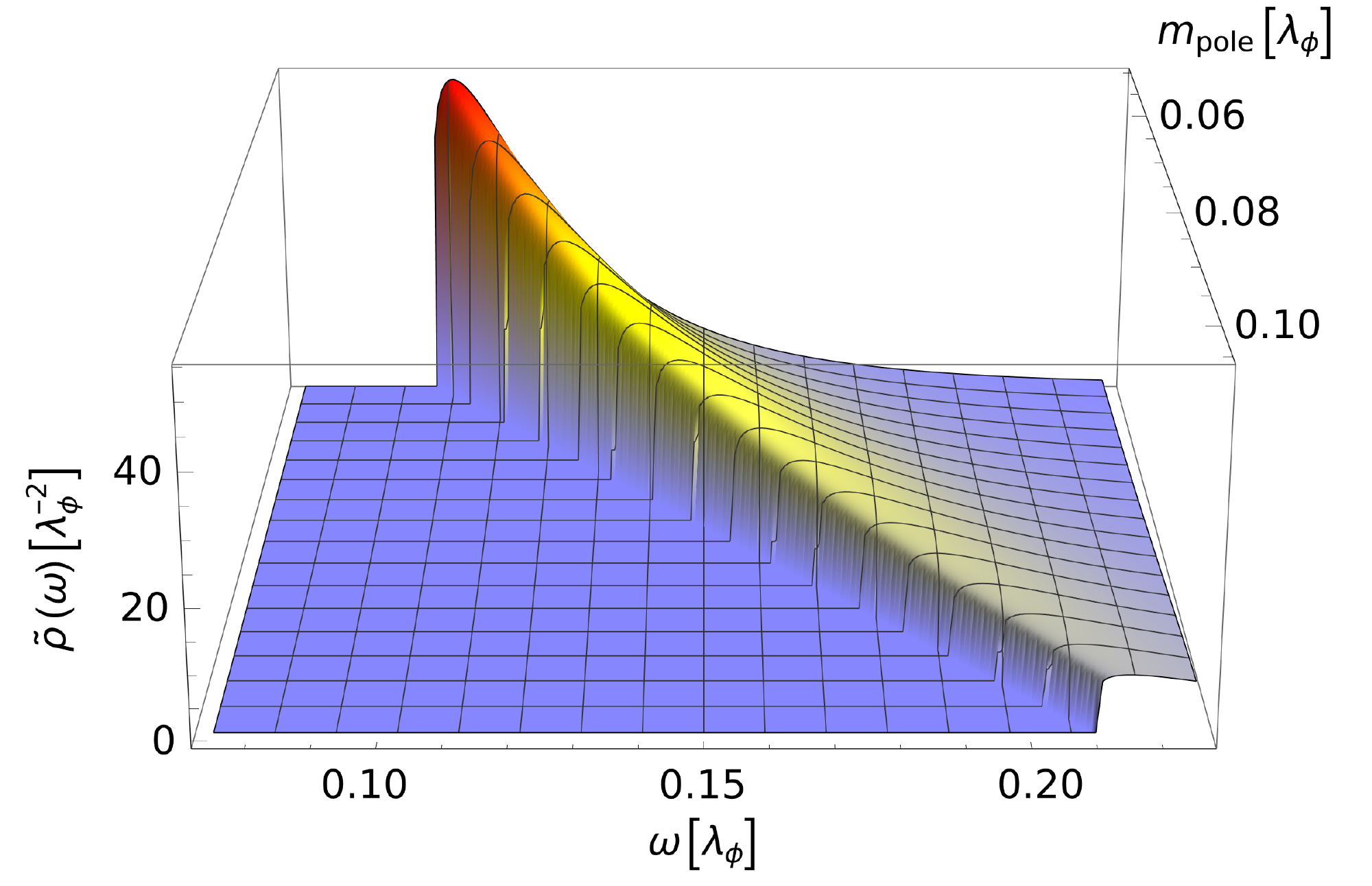}
	\caption{Scattering tail $\tilde{\rho}_k$ for vanishing spatial momentum $\vec p=0$ as a function of the spectral value $\omega$ and the pole mass $m_\tinytext{pole}$ for $1/20\leq m_{\tinytext{pole}}/\lambda_{\phi}\leq 1/10$. \hspace*{\fill}} 
	\label{fig:3dplot_tail}
	\end{subfigure}
	\hspace*{.51cm}
	\begin{subfigure}{.48\linewidth}
	\centering\vspace*{2.5mm}
		\includegraphics[width=.9\textwidth]{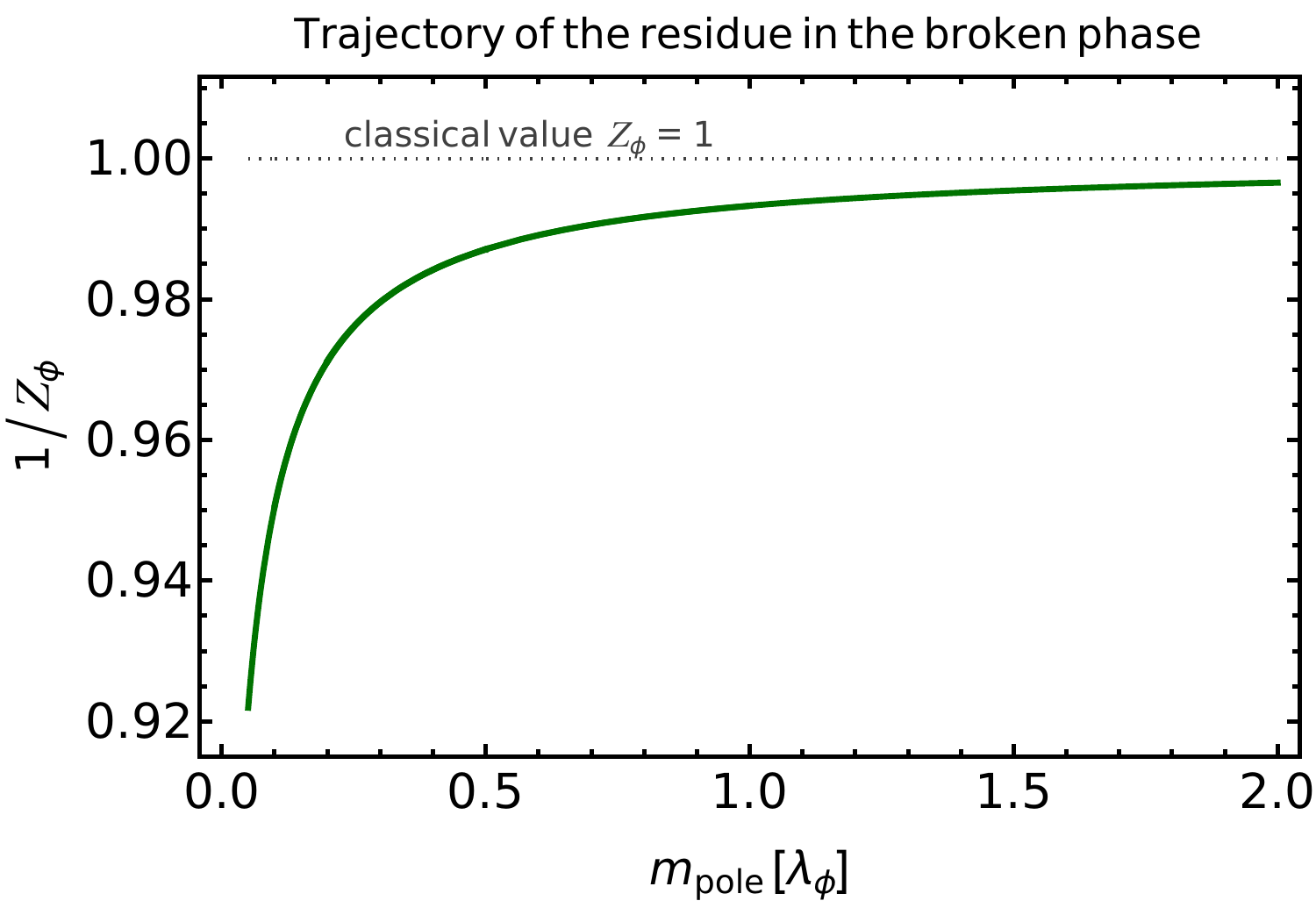}
		\vspace{2.5mm}\caption{Amplitude ${1}/{Z_\phi}$ of the pole contribution of the spectral function \labelcref{eq:scale-dep-spec} as a function of the pole mass $m_\tinytext{pole}$ for $1/20\leq m_{\tinytext{pole}}/\lambda_{\phi}\leq 1/2$. The classical value for $Z_\phi$ is indicated in grey. \hspace*{\fill}}
		\label{fig:3dplot_Z}
	\end{subfigure}
	\caption{Spectral function $\rho$, \labelcref{eq:scale-dep-spec}, for different pole masses $m_{\tinytext{pole}}/\lambda_\phi$, measured in the fixed coupling $\lambda_\phi$. }
\label{fig:3dplot}
\end{figure*}

The spectral structure of the diagrams allows for a simple discussion of the emergent scattering thresholds that can be easily tracked within spectral functional approaches. An illustrative example is given by the contribution of the vacuum polarisation diagram to the spectral function of a single scalar field: It features a branch cut that opens at the sum of the spectral masses of the two propagators. The spectral function entering the diagram consists of a mass pole at $m_\tinytext{pole}$ and a sum of scattering continua $\rho_N$ starting at $N m_\tinytext{pole}$ with $N \geq 2$. It follows straightforwardly from the analytic structure of that diagram that substituting scattering contributions $\rho_N$ and $\rho_M$ for the two internal lines directly yields a contribution to $\rho_{N+M}$. This demonstrates how any scattering structure, once seeded, gives rise to higher scattering contributions.

\section{Spectral functional renormalisation group}
\label{sec:frg}

In the spectral fRG approach put-forward in \cite{Fehre:2021eob, Braun:2022mgx}, the quantum effective action of the theory at hand is obtained by starting with a theory with an asymptotically large classical pole mass $m_\phi \to \infty$, and then lowering the mass successively until the physical point is reached. The respective classical action is given by \labelcref{eq:S} with 
\begin{align}
	S[\phi] &= \int\! \mathrm{d}^3x \,\bigg\{ \frac{1}{2} \phi\Bigl(-\partial^2+ Z_\phi \,\mu \Bigr)\phi + \frac{\lambda_\phi}{4!} \phi^4\bigg\}\,, 
 \label{eq:scale-dep-S}
\end{align} 
with positive or negative $\mu$. The wave function $Z_\phi$ has been introduced for convenience, anticipating the emergence of a wave function. For asymptotically large pole masses we have $Z_\phi\to 1$, see \Cref{fig:3dplot_Z}. Then, \labelcref{eq:scale-dep-S} reduces to \labelcref{eq:S}, and the pole mass is given by 
\begin{align} 
	 m_\phi^2 = \mu -3 \mu\, \theta(-\mu)\,, 
\label{eq:ClassicalPole}
\end{align}
capturing both the symmetric and broken phase. This setup captures both, theories deep in the symmetric phase with $\mu\to +\infty$ and theories deep in the broken phase with $\mu\to -\infty$.

\subsection{Functional Callan-Symanzik equation}
\label{sec:funCS}

The infinitesimal change of the quantum effective action $\Gamma[\phi]$ under a change of the mass $\mu$ is governed by the manifestly finite \textit{renormalised} Callan-Symanzik equation, \cite{Fehre:2021eob, Braun:2022mgx}, 
\begin{subequations}
	\label{eq:CSflow-gamma}
\begin{align}\nonumber 
\mu\partial_\mu \Gamma[\phi] =&\,  \frac12\,  \left(1- \frac{\eta_\phi}{2}\right)Z_\phi \mu\, \text{Tr} \,\Bigl[ \,G[\phi]+\phi^2\Bigr] \\[1ex] 
&\hspace{3cm}- \frac12 \, \mu \partial_\mu   S_{\text{ct}}[\phi] \,, 
\label{eq:CSflow-gamma-fRG}
\end{align}
with the anomalous dimension
\begin{align}
	\eta_\phi= -2 \frac{\mu \partial_\mu \, Z_\phi}{Z_\phi}\,,
\label{eq:etaphimu}
\end{align}
The factor 2 in \labelcref{eq:etaphimu} takes into account that $\mu$ has mass dimension 2. The argument $\phi$ in \labelcref{eq:CSflow-gamma} is the mean field. The term $\mu\partial_\mu S_{\text{ct}}[\phi] $ in the second line of~\labelcref{eq:CSflow-gamma-fRG} is the flow of the counter terms that renders the flow equation finite, where the factor $1/2$ was added for convenience. The loop term on the right-hand side depends on the full field-dependent propagator. In the momentum basis it is given by 
\begin{align}
	G[\phi](p,q) = &\,\langle \varphi \varphi \rangle_c(p,q)  = \frac{1}{\Gamma^{(2)}[\phi]}(p,q) \,,
\label{eq:CSflow-gamma-Prop}
\end{align}
where the subscript ${}_c$ refers to the connected part of the two-point function and the mean field $\phi$ is given by the expectation value of the quantum field $\varphi$, i.e., $\phi=\langle \varphi\rangle$.
\end{subequations}
In the momentum basis, the trace in \labelcref{eq:CSflow-gamma-fRG} corresponds to a momentum integral. 
\begin{figure*}
\centering
	\includegraphics[width=.99\linewidth]{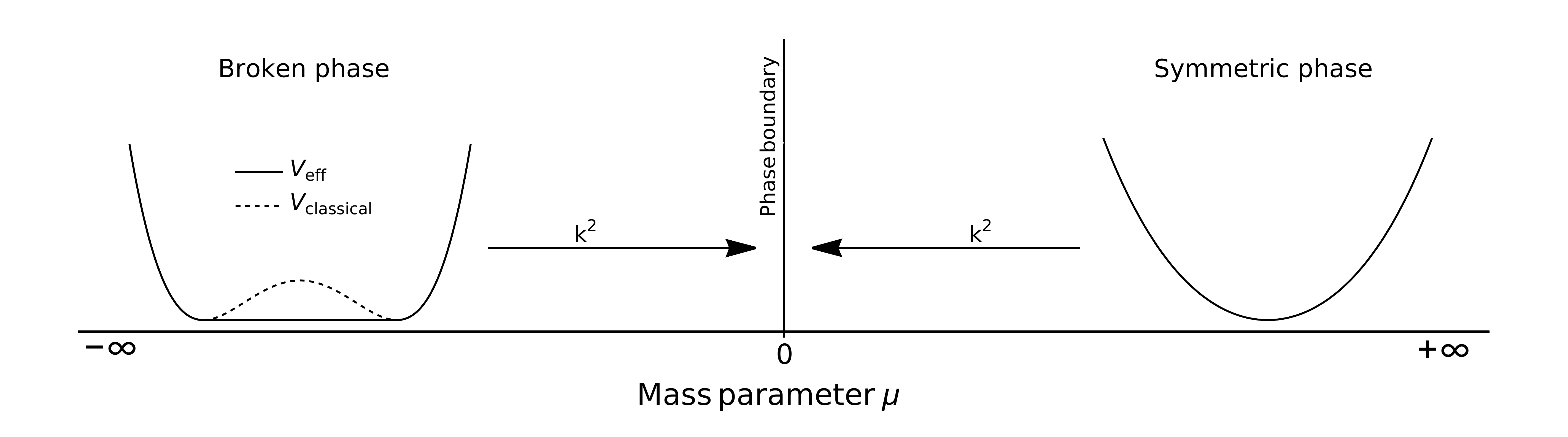}
	\caption{Schematic phase diagram with respect to the mass-parameter $\mu$. The phase boundary is located at $\mu=0$. The flow is initiated in the deep UV, \ie $|\mu|=k^2 \to \infty$ with the respective (classical) initial effective potential. \hspace*{\fill}}
	\label{fig:FlowLeftRight}
\end{figure*}
Note that the effective action $\Gamma[\phi]$ in \labelcref{eq:CSflow-gamma} includes the full mass term $1/2 \int_x \mu\,\phi^2$ in contradistinction to the effective action used in standard fRG momentum-shell flows. There, the momentum dependent regulator part of the mass term is subtracted, and the physical theory is reached when it vanishes. In the present setup, the $\mu$-dependent effective action is that of a physical theory with mass parameter $\mu$, and the flow is one in (physical) theory space. In contrast to usual momentum shell flows, this physical flow is both manifestly Lorentz invariant and sustains causality of physical correlation functions throughout the flow, allowing for the use of the Källén-Lehmann spectral representation \labelcref{eq:specrep}.

Compared to the Wetterich equation,~\cite{Wetterich:1992yh}, with a momentum-dependent infrared regularisation, the novel ingredient in the functional Callan-Symanzik equation~\labelcref{eq:CSflow-gamma} is the explicit counter term flow $\mu\partial_\mu S_{\text{ct}}[\phi]$. The counter term flow has been derived in a manifestly finite limit of standard momentum cutoff flow equations as discussed in detail in~\cite{Braun:2022mgx}. The derivation entails that the flow of the counter term originates from a closed one-loop expression such as the trace in~\labelcref{eq:CSflow-gamma} itself, i.e., $\mu \partial_\mu S_{\text{ct}}[\phi] \sim \text{diagramms}$. Accordingly, the counter term flow contains no tree-level contributions to the respective correlation functions. This entails that classical values of the correlation functions are solely given by the respective choice of tree-level values specified in the classical action~\labelcref{eq:scale-dep-S}, and in particular cannot be further changed by specification of renormalisation conditions. In consequence, the latter can only be used to renormalise the flow contributions, but not the initial conditions of the flow. This excludes, for example, that the counter term flow rearranges the theory from the symmetric into the broken phase or vice versa by $\mu\partial_\mu  S_\textrm{ct} \propto \pm \textrm{const.}\, \mu\int  \phi^2$. However, the counter term can contain similar terms proportional to $\lambda_\phi/m_{\tinytext{pole}} = \lambda_\text{eff}$.

In particular, the counter term flow allows for \textit{flowing renormalisation conditions}, and we shall use it to adjust a \textit{flowing on-shell renormalisation}, based on the spectral on-shell renormalisation put forward in  \cite{Horak:2020eng}. Then, the pole mass $m_\tinytext{pole}$ is identified with $m_\phi$ in \labelcref{eq:ClassicalPole} in both phases, $m_\tinytext{pole}^2= \mu -3 \mu\, \theta(-\mu)$. In this physical RG scheme, the phase transition between the symmetric and broken phase happens for $m^2_\phi=0$. Hence, we  approach the phase transition both from the broken and the symmetric phase in the limit $\mu\to 0$, and the flows are taking place in the one or the other phase, see \Cref{fig:FlowLeftRight}. Thereby, our setup avoids flows through the strongly interacting phase transition regime, which are usually present in momentum cutoff flows. This minimises the systematic error stemming from the strong dynamics in the vicinity of a phase transition, where the flows are highly sensitive to truncation artefacts. However, it is in principle possible to flow through the phase transition, what can be advantageous if it is difficult to identify a proper starting point in one phase, at which the theory is particularly simple. An example for this situation can be found in quantum mechanics, where the theory for $\mu \to -\infty$ does not approach the classical limit but the instanton dominated regime.

In the present work, we consider the flow of the inverse propagator within the spectral representation. The flow is given by 
\begin{align} \nonumber
\mu	 \partial_\mu \Gamma^{(2)}(p^2) =&\,  \left(1-\frac{\eta_\phi}{2} \right)  Z_\phi  \mu\, \left[ D_\tinytext{pol}(p^2)- \frac12 D_\tinytext{tad}(p^2)\right]\\[1ex]
	& + \left(1-\frac{\eta_\phi}{2} \right) Z_\phi \mu -\frac12 \,\mu\partial_\mu  S^{(2)}_{\text{ct}}, 
\label{eq:CSflowInverseProp}
\end{align}
where $D_\text{tad}$ and $D_\text{pol}$ refer to the tadpole and polarisation diagram, see~\Cref{fig:2pointflow}. Their general form, in terms of the spectral representation for the propagator and four-point function, is discussed in~\Cref{sec:specDiags}. Moreover, all quantities in \labelcref{eq:CSflowInverseProp} depend on the chosen background $\phi$. For general spacetime dependent backgrounds $\phi(x)$ this would lead to $\Gamma^{(2)}[\phi](p,q)$. In the explicit computations we consider the background $\phi_0$, which is the constant solution of the equation of motion
\begin{align}
\left.	\frac{\delta\Gamma[\phi]}{\delta \phi}\right|_{\phi=\phi_0}=0\,. 
	\label{eq:EoM} 
\end{align}
With this physical choice for the background, the general field-dependent propagator~\labelcref{eq:CSflow-gamma-Prop} reduces to the physical propagator $G(p^2)$ in the absence of source terms, 
\begin{align} 
	G(p^2) =  \frac{1}{\Gamma^{(2)}[\phi_0](p^2)}\,.
\label{eq:prop}
\end{align}
In the symmetric phase, we have $\phi_0=0$, while $\phi_0\neq 0$ signals the broken phase. At constant fields the propagator \labelcref{eq:CSflow-gamma-Prop} reduces to $G(p,q)= G(p^2) (2\pi)^d\delta(p+q)$. Similarly we have $\Gamma^{(2)}[\phi_0](p,q) = \Gamma^{(2)}[\phi_0](p^2)(2\pi)^d\delta(p+q)$. 

In three dimensions the two phases are separated by a second order phase transition in the Ising universality class. From now on we drop the field argument $\phi_0$. It is implicitly understood that all correlation functions are evaluated at $\phi=\phi_0$. 

In a final step, we substitute $\mu$ with $\pm k^2$, to keep the relations to standard fRG flows with momentum cutoffs simple, where $k$ is commonly used. This facilitates the comparison and benchmarking of the real-time results obtained with the spectral fRG. For example, the three dimensional $\phi^4$-theory has been studied abundantly within the Euclidean fRG, including systematic studies of the convergence of approximation schemes, for a recent review see \cite{Dupuis:2020fhh}. These results carry over straightforwardly to the present approach, and the Euclidean correlation functions obtained from the spectral functions can be directly compared. This substitution leads us to 
\begin{align}
k^2=|\mu|\, , \qquad \partial_t=k\,\partial_k = 2 \mu \partial_\mu  \,\,,
\label{eq:kdef}
\end{align}
where the (negative) RG-time $t=\log(k/k_\text{ref})$ is measured relatively to a suitable reference scale or mass.

\subsection{Spectral on-shell renormalisation}
\label{sec:SpecSetup}

\begin{figure}[t] 
\centering
	\begin{minipage}{0.99\linewidth}
		\centering 
	\includegraphics[width=\linewidth]{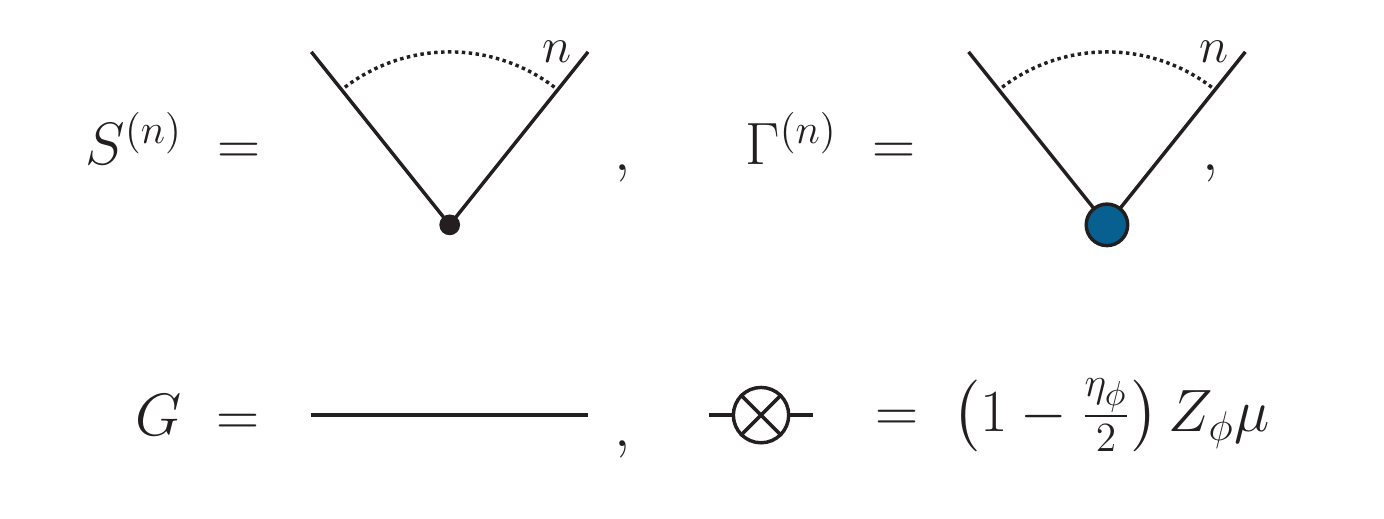}
	\caption{Diagrammatic notation used throughout this work:
	Lines stand for full propagators, small black dots stand for
	classical vertices, and larger blue dots stand for full vertices. The crossed circle 		represents the scale derivative of the mass parameter. \hspace*{\fill}}
	\label{fig:Notation}
	\end{minipage}
\end{figure}

We proceed with discussing the on-shell spectral renormalisation, using the direct access to Minkowskian momenta. In \mbox{(1+2)-dimensions}, both diagrams in the CS flow \labelcref{eq:CSflowInverseProp} are manifestly finite, and the flow of the counter term action $\mu\partial_\mu S_\text{ct}$ only guarantees the implementation of the chosen renormalisation conditions. The \mbox{(1+2)-dimensional} $\phi^4$-theory is super-renormalisable, and one only has the renormalisation condition for the mass. Now we use on-shell renormalisation to keep the full pole mass on the classical input mass  \labelcref{eq:ClassicalPole} with $m_\tinytext{pole}^2=k^2$ in the symmetric phase, and $m_\tinytext{pole}^2=2k^2$ in the broken phase. This leads us to 
\begin{enumerate}[label=(\roman*)]
\item symmetric phase: 
\begin{align} 
	\Gamma^{(2)}[\phi_0] \Big|_{p^2=-k^2}=0\,, 
\label{eq:onshellrenSymmetric}
\end{align}
\item broken phase:
\begin{align} 
	\Gamma^{(2)}[\phi_0] \Big|_{p^2=-2k^2}=0\,. 
\label{eq:onshellrenBroken}
\end{align}
\end{enumerate}
In the symmetric phase, the first allowed scattering process is the $1\to 3$ scattering, and the onset of the scattering continuum is located at three times the pole mass. In turn, in the broken phase with $1\rightarrow 2$ scattering, the onset of the scattering continuum of the spectral function is located at twice the pole mass. Thus, the spectral function \labelcref{eq:scale-dep-specIntro} reads 
\begin{align} 
	\rho(\lambda) = \frac{2\pi}{Z_\phi} \delta (\lambda^2 - m_\tinytext{pole}^2) + \theta (\lambda^2-m^2_\tinytext{scat} ) \tilde \rho(\lambda) \,,
\label{eq:scale-dep-spec}
\end{align}
with $m_\tinytext{scat}=3m_\tinytext{pole}$ (symmetric phase) and $m_\tinytext{scat}=2 m_\tinytext{pole}$ (broken phase). In~\Cref{fig:3dplot}, we show the scale evolution of the spectral function $\rho$ in the broken phase: in~\Cref{fig:3dplot_tail} we depict the scattering tail $\tilde \rho$, and in~\Cref{fig:3dplot_Z} we depict the amplitude of the pole contribution. All quantities are measured relative to the coupling $\lambda_\phi$. 
\begin{figure}
	\centering
\begin{minipage}{0.99\linewidth}
	\centering
	\includegraphics[width=\linewidth]{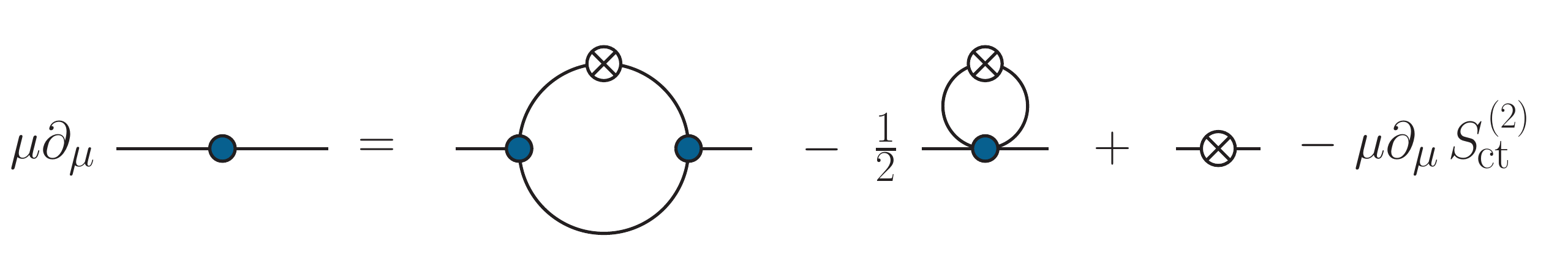}
	\caption{Renormalised CS equation for the inverse propagator. The notation is given in \Cref{fig:Notation}. \hspace*{\fill}}
	\label{fig:2pointflow}
\end{minipage}
\end{figure}
The spectral tail is rising towards smaller pole masses for a fixed classical coupling, and in turn the amplitude $1/Z_\phi$ of the pole contribution is decreasing. In combination the sum rule \labelcref{eq:specnorm} holds during the evolution. The growing importance of the scattering processes can be understood from the fact that the dynamics of the theory only depend on the dimensionless ratio $\lambda_{\phi}/m_\phi$ with $m_\phi\propto k$. Hence, the effective coupling grows strong for smaller pole masses and on the other hand the dynamics of the theory are vanishing for asymptotically large pole masses.

In contrast to the Callan-Symanzik or mass regulator used in the present work, commonly used regulators in Euclidean flows decay for momenta larger than the IR cutoff $k$. This provides manifestly finite flows without the need of further renormalisation. Moreover, for Euclidean momenta, the respective flows of lower order correlation functions decay faster than for a CS regulator. In Minkowski space, however, the CS or mass regulator has the welcoming property, that the one-loop flow of $\rho(\omega)$ contains only classical correlation functions and is maximally local. While this is trivial in the symmetric phase where the one-loop flow only shifts the pole mass and does not generate a scattering continuum, it is non-trivial in the broken phase. There, the flow of the scattering continuum is given by a single delta function at the onset of the scattering spectrum, which originates from $\partial_t \, \text{Im}\, \Gamma^{(2)}\propto \delta(\omega^2 - 4 m_\tinytext{pole}^2)$. Since the mass pole constitutes the dominant part of the propagator, the flow of the spectral function at spectral values larger than the flowing onset $2m_\tinytext{pole}$, which is solely induced by the scattering tail, is sub-leading.

\subsection{Flowing with the minimum}
\label{sec:flowingMinimum}

\begin{figure*}[ht]
	\centering
	\includegraphics[width=.95\linewidth]{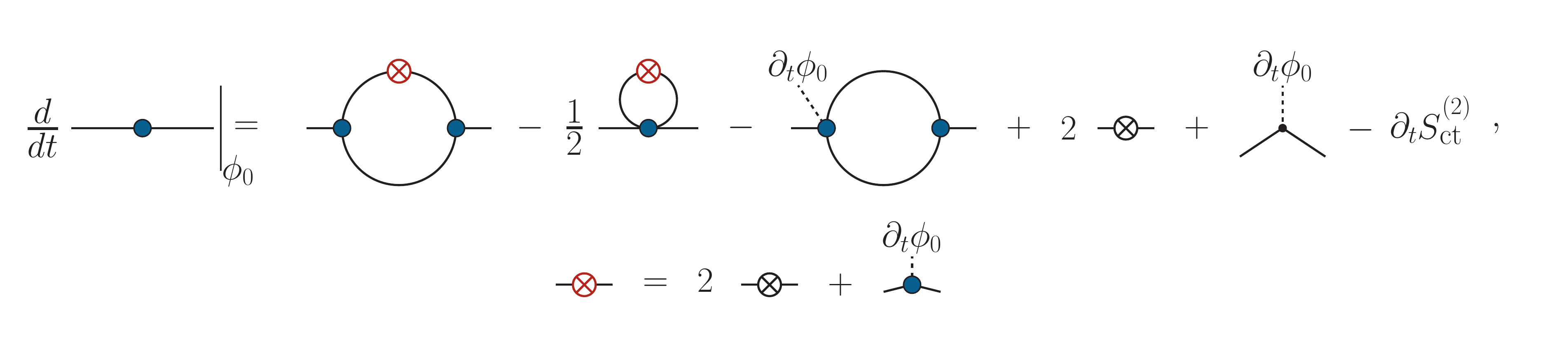}
	\caption{Diagrammatic representation of the flow of the two-point function on the flowing minimum in the broken phase. The notation is given in \Cref{fig:Notation}. The red crossed circle comprises the scale derivative of the mass parameter and the three-point function, where the additional factor of $2$ comes from the change from $\mu$ to $k$. The dashed lines indicate the contraction with $\partial_t\phi_0$.  \hspace*{\fill}
		}
	\label{Diag:flowingMinimum2Point}
\end{figure*}
In general, the flow equation \labelcref{eq:CSflow-gamma-fRG} can be evaluated for arbitrary values of the external field $\phi$, which requires the inclusion of the full effective potential. However, this goes beyond the scope of this work, and we simply evaluate the flow on the solution $\phi_0$ of the equation of motion \labelcref{eq:EoM}. This is a commonly used truncation as it gives access to the physical correlation functions. 

In the present fRG approach with the spectral CS regulator, the flow takes place in theory space and the effective action is physical for all values of $k$. In the broken phase, the minimum of the full effective potential depends on $k$, and the total mass flow of the two-point function is given by the flow diagrams originating from the CS equation, $ \partial_t  \Gamma^{(2)}[\phi_0](p)$ and a term proportional to the mass flow of $\phi_0$, 
\begin{align}
	\frac{d}{dt}\Gamma^{(2)}[\phi_0](p) = \partial_t \Gamma^{(2)}[\phi_0](p) + \left(\partial_t \phi_0 \; \Gamma^{(3)}[\phi_0]\right)(p)\,.
\label{eq:FullFlowGamma2}
\end{align}
The novel ingredient in the present setup originates in the tree-level $k$-dependence of $\phi_0 \approx \sqrt{6 k^2/\lambda_\phi} + \mathcal{O}(\lambda_\phi k)$, where the second term comprises the loop corrections. This tree-level dependence is usually absent in the flow of the minimum in standard momentum-shell flows. There, $\partial_t \phi_0$ only comprises the effects of the momentum shell integration and hence is inherently one-loop and beyond. The tree-level $k$-dependence of $\phi_0$ in the present case triggers a tree-level $k$-dependence of $\partial_t\phi_0 \Gamma^{(3)}(p)$ and the tree-level flow of the physical two-point function considered here reads
\begin{align}\label{eq:treelevelflow}
\frac{d}{dt}\,\Gamma^{(2)} \Big|_{\tinytext{tree-level}} &= -2k^2 + \partial_t \phi_0 S^{(3)}[\phi_0] = 4k^2 \,,
\end{align}
where the classical three-point function is given by $S^{(3)}[\phi]=\lambda_\phi \phi$. Note that only the combination of both terms leads to the expected positive flow of the physical mass, while the flow of the mass parameter $-k^2$ has a negative sign. 

\begin{figure}[t]
	\centering
	\includegraphics[width=.99\linewidth]{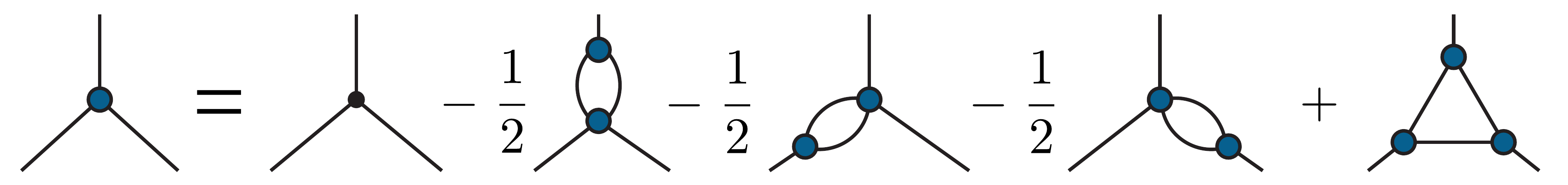}
	\caption{Truncated DSE for the three-point function in the skeleton expansion. The notation is given in \Cref{fig:Notation}. \hspace*{\fill}}
	\label{Diag:threepoint}
\end{figure}
To obtain the full momentum structure of the second term in of \labelcref{eq:FullFlowGamma2}, we first note that the additional leg of the three-point function is always augmented with an incoming momentum of zero, as it is contracted with the scale derivative of a constant field. The full momentum dependence can then be incorporated via the DSE of the three-point function, which allows for an exact diagrammatic flow of the two-point function on the physical minimum. In the presence of approximations, a fully self-consistent treatment would require us to use the integrated flow of $\Gamma^{(3)}[\phi_0](p,0)$. However, also the flow of $\Gamma^{(3)}[\phi_0]$ includes a similar additional term as in~\labelcref{eq:treelevelflow}, which is proportional to the four-point function. 
To avoid solving the flow of the three- and four-point function, we resort to the DSE to include the leading momentum dependence of the (contracted) three-point function. To ensure the correct RG-scaling of the flow equation, we further employ the skeleton expansion in the DSE, where every vertex is dressed. 
Approximating $\Gamma^{(n>4)}\approx 0 $ and dropping the remaining two-loop diagrams, we arrive at the simple diagrammatic structure of the three-point function depicted in \Cref{Diag:threepoint}. 

Additionally, using the DSE for $\Gamma^{(3)}$ in \labelcref{eq:FullFlowGamma2} demonstrates the structure of the flow as a total derivative. To make this explicit, we choose the vertical leg in \Cref{Diag:threepoint} to be contracted with $\partial_t \phi_0$. Then, the three-point functions connected to this leg carry only internal momenta, and we approximate them as constant. With that, the first fish- and the triangle diagram in \Cref{Diag:threepoint} are proportional to the tadpole and polarisation diagram respectively, and the second term on the RHS of the flow~\labelcref{eq:FullFlowGamma2} reads 
\begin{align}\label{eq:threepoint} \nonumber
 \hspace{-2mm} \Big(\partial_t \phi_0 \; \Gamma^{(3)}[\phi_0]\Big) (p) = &\, \partial_t \phi_0\, \bigg(S^{(3)}[\phi_0] - \frac12 \Gamma^{(3)}[\phi_0] \,D_\tinytext{tad}(p)  \\[1ex] 
 & \hspace{-3mm} - D_\tinytext{fish}(p) + \Gamma^{(3)}[\phi_0] \,D_\tinytext{pol}(p)\bigg)\,.
\end{align}
Note that the explicit three-point functions on the right-hand side are now momentum independent. For the full expressions of the spectral diagrams we refer to \Cref{sec:specDiags}. We discuss our approximations for the remaining vertices in \Cref{sec:phi4}.

Substituting \labelcref{eq:CSflowInverseProp} and \labelcref{eq:threepoint} into~\labelcref{eq:FullFlowGamma2}, we eventually arrive at the full flow equation of the two-point function. Its diagrammatic representation is depicted in \Cref{Diag:flowingMinimum2Point}. It is illuminating to consider the one-loop structure of the flow, where the nature of the flow being a total derivative can be read off \Cref{Diag:flowingMinimum2Point}. Then, the red crossed circle comprises the total derivative of the internal propagators in the (one-loop) polarisation and tadpole diagram, while the fish-diagram accounts for the running of the three-point vertices. The full equation reads
\begin{subequations}
\begin{align}\nonumber 
	&\hspace*{-.14cm}\frac{d}{dt}\Gamma^{(2)}[\phi_0](p)=\;\left(\partial_t\phi_0\right)S^{(3)}[\phi_0] -\left(2-\eta_\phi\right)Z_\phi k^2\\[1ex]
	& \hspace{.5 cm} +\dot{\mathcal{R}}  \left[-\frac12 D_\tinytext{tad} + D_\tinytext{pol}\right]  -\partial_t \phi_0 D_{\tinytext{fish}}-\partial_t S_\tinytext{ct}[\phi_0] \, ,
	\label{eq:FullFlowGammaFinal1}
\end{align}
where
\begin{align} \label{eq:rdot}
\dot{\mathcal{R}}=\left(\partial_t \phi_0 \Gamma^{(3)}[\phi_0] - \left(2-\eta_\phi \right) Z_\phi k^2 \right)\,,
\end{align}
\end{subequations}
is represented as red crossed circle in \Cref{Diag:flowingMinimum2Point}. Note the appearance of a relative minus sign in front of the mass derivative contribution (second term) to~\labelcref{eq:rdot} due to \mbox{$\mu = -k^2$} in the broken phase. The first line in \labelcref{eq:FullFlowGammaFinal1} carries the trivial, tree-level running of inverse propagator. It consists of the running of the mass parameter and the classical part of the three-point function, connected to the flow of the minimum. Its mean-field value cannot be altered by the renormalisation condition and is, analogue to the respective term in $\dot{\mathcal{R}}$, crucial to recover the correct sign of the flow, see \labelcref{eq:treelevelflow}. A detailed evaluation of \labelcref{eq:FullFlowGammaFinal1} can be found in \Cref{sec:FlowBroken}.

\section{Approximations and real-time flows in the symmetric and broken phase}
\label{sec:phi4}

In the following section, we discuss the approximations used for the higher correlation functions, which lead to non-trivial spectral flow-equations in both phases. This enables us to write down the renormalised flow equations for the two-point function and evaluate them on the real frequency axes. 

In the $\phi^4$-theory, correlation functions of an odd number of fields, $\Gamma^{(2n+1)}[\phi]$, are proportional to the mean field $\phi$. In the present approximation we only consider three- and four-point functions, setting all the higher correlation functions to zero:
\begin{align}
	\Gamma^{(n>4)}\approx 0\,.
	\label{eq:Approxn>4}
\end{align}
Then, the three-point function is proportional to a product of the four-point function and $\phi_0$. This closes our approximation. 

\begin{figure}[t]
	\centering
	\includegraphics[width=.99\linewidth]{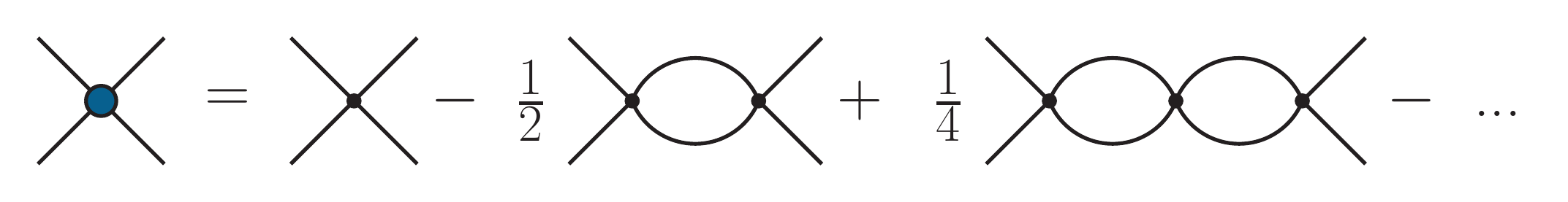}
	\caption{Bubble resummed four-point function. The notation is given in \Cref{fig:Notation}. \hspace*{\fill}}
	\label{fig:bubble}
\end{figure}

For constant vertices, the tadpole diagram only provides a constant contribution to the flow of the two-point function. This contribution is absorbed completely in the on-shell renormalisation condition \labelcref{eq:onshellrenBroken} and \labelcref{eq:onshellrenSymmetric}, for the broken and symmetric phase respectively. In the symmetric phase of the theory with $\phi_0=0$, the tadpole is the only contribution to the flow of the two-point function.
Hence, the scattering tail originates only from the non-trivial momentum dependence of the four-point function. In a first but important step towards the full momentum dependence of $\Gamma^{(4)}(p_1,...,p_4)$ we use an $s$-channel resummation of the full four-point function, see~\Cref{fig:bubble},
\begin{align} 
	\Gamma^{(4)}(p^2) = \frac{\lambda_{\phi}}{1+\frac{\lambda_{\phi}}{2} \int_q G(p+q)G(q)} \, .
\label{eq:resummed-four-pt-func}
\end{align}
In \labelcref{eq:resummed-four-pt-func}, $p^2=s=(p_1+p_2)^2$ is the $s$-channel momentum, and we choose vanishing t and u channels to perform the resummation: $(p_3-p_1)^2 = (p_1 - p_4)^2= 0 $. This approximation admits the simple spectral representation~\labelcref{eq:specrep4point} of the four-point function, see also~\cite{Horak:2020eng}. 

We emphasise that~\labelcref{eq:resummed-four-pt-func} only holds true in the symmetric phase. In contrast, in the broken phase the flow or BSE for the four-point function contains additional diagrams with two or four three-point vertices. Their combined contributions are readily estimated and are suppressed by a factor $1/8$. Hence, they are dropped in the following computation. Accordingly, we use~\labelcref{eq:resummed-four-pt-func} in both phases. 

Note also, that the four-point function exhibits a bound state pole below $2 m_\tinytext{pole}$ close to the phase transition. This is discussed for example in~\cite{Caselle:2001im} in terms of a Bethe-Salpeter equation, and indeed seen in lattice and fRG calculations, see~\cite{Agostini:1996xy, Caselle:1999tm, Rose:2016wqz}. The present $s$-channel resummation for the four-vertex does not include the resonant channel. A full bound state analysis and the systematic inclusion of other channels will be considered elsewhere.

It is left to specify the three-point function $\Gamma^{(3)}(p_1,p_2,p_3)$ in \labelcref{eq:FullFlowGammaFinal1}. In contrast to the pivotal importance of the momentum dependence of the four-point function, that of the three-point function is averaged out in the vacuum polarisation and the fish diagram. For the sake of simplicity, we therefore approximate the full vertex by its value at vanishing momenta, $p_i=0$ for $i=1,2,3$. There, the three-point function is given by the third derivative of the effective potential on the equations of motion, $V^{(3)}_\textrm{eff}(\phi_0)$. The effective potential $V_\textrm{eff}(\phi)$ is the quantum analogue of the classical potential, and is nothing but the effective action $\Gamma[\phi]$, evaluated for constant fields $\phi_c$, 
\begin{align}
	 V_\textrm{eff}(\phi_c) = \frac{1}{{\cal V}} \Gamma[\phi_c]\,,\qquad {\cal V} =\int d^3 x\,. 
\label{eq:DefofVeff}
\end{align}
Due to the $Z_2$-symmetry of the $\phi^4$-theory under $\phi\to -\phi$, the effective potential is symmetric, $V_\textrm{eff}(-\phi) = V_\textrm{eff}(\phi)$. Moreover, it admits an expansion about the solution to the equation of motion, $\phi^2=\phi_0^2$, which is valid for $\phi^2\geq \phi_0^2$. The latter constraint on the modulus of $(\phi^2 -\phi_0^2)$ originates from the fact that the classical effective potential is the double Legendre transform of the classical potential. In the case of a non-convex potential it is simply the convex hull. Schematically, this is depicted in \Cref{fig:FlowLeftRight}. 

The effective potential satisfies its own flow equation, and for the sake of completeness we briefly discuss its derivation and explicit form in \Cref{sec:CS-EffPot}, more details can be found in \cite{Braun:2022mgx}. The present computation can be augmented by the full flow of the effective potential, effectively leading to a cutoff dependence of the coupling $\lambda_\phi$ in \labelcref{eq:resummed-four-pt-func} and similar changes. While this provides further quantitative precision to the computation, it goes beyond the scope of the present work and will be presented elsewhere. Here we shall consider the expansion up to $(\phi^2-\phi_0^2)^2$, dropping higher order terms in accordance with \labelcref{eq:Approxn>4}, and discuss the symmetric and broken phase separately in \Cref{sec:BrokenPhase} and \Cref{sec:SymPhase} below.

\subsection{Symmetric phase}
\label{sec:SymPhase}

In the symmetric phase with $\phi_0=0$ in \labelcref{eq:EoM}, we use a Taylor expansion about $\phi^2=0$ for the effective potential, 
\begin{align}
	V_\textrm{eff}(\phi)=   \sum_{n=1}^\infty \frac{\lambda_{n}}{2 n!} \phi^{2n}\,. 
	\label{eq:Veffsym}
\end{align}
The first two couplings, $\lambda_1$ and $\lambda_2$ are related to the correlation functions $\Gamma^{(2)}$ and $\Gamma^{(4)}$ considered here. Hence, the coupling $\lambda_1$ agrees with the curvature mass squared in the symmetric phase, where the curvature mass is defined as
\begin{align}\label{eq:mcurv}
m_\tinytext{curv}^2=V_\textrm{eff}^{(2)}(\phi_0)=\Gamma^{(2)}[\phi_0](p=0)\,, 
\end{align}
in both phases. Moreover, the coupling $\lambda_2$ is nothing but the full four-point function, evaluated at vanishing momentum. In summary we have    
\begin{align}
\lambda_1=m_\tinytext{curv}^2\,,\qquad 	\lambda_2= \Gamma^{(4)}(p=0) \,.
	\label{eq:lambdaphi-4}
\end{align}
For the initial UV pole mass  $m_\tinytext{pole}=\Lambda$, the curvature mass and the pole mass agree, $\lambda_1=\Lambda^2$, and the initial coupling is the classical one, $\lambda_2=\lambda_\phi$. Hence, the initial effective potential $V_\tinytext{UV}(\phi)$ at $k=\Lambda$ reads 
\begin{align}
	V_\tinytext{UV}(\phi) = \frac{1}{2} \Lambda^2 \phi^2 + \frac{1}{4!} \lambda_\phi\, \phi^4\,. 
	\label{eq:VeffUVsymmetric}
\end{align}
With the above approximations, all higher correlation functions are fixed and the flow equation of the two-point function on the real frequency axes reads 
\begin{align}
	\partial_t\Gamma^{(2)}(\omega_+)&=- \frac{Z_\phi\left(2-\eta_\phi\right)k^2}{2} D^{\text{dyn}}_{\text{tad}}(\omega_+) + 2k^2 - \partial_t \hat{S}_{\text{ct}}^{(2)},
	\label{eq:symmetricflow}
\end{align}
where the retarded limit is given by $\omega_+= -\imag (w+\imag 0^+)$ and is explicitly carried out in~\Cref{sec:analyticintegrands}. $\hat{S}^{(2)}_{\text{ct}}$ is given schematically by 
\begin{align}
\hat{S}^{(2)}_{\text{ct}} = \text{diagramms}\left(p^2=-k^2\right) \,.
\end{align}
We denoted the counterterm action with a tilde since we already dropped constant terms in the flow of order $\lambda_\phi k$. Hence, only the dynamic part of the tadpole $ D^{\tinytext{dyn}}_\tinytext{tad}$ contributes. It arises from the scattering tail of $\Gamma^{(4)}(p)$ and carries the spectral structure of the polarisation diagram, see \labelcref{eq:tadpole}. In particular, the deviation of the constant term in \Cref{fig:2pointflow} from its classical value, $2k^2$, is of order $(\lambda_\phi k)$ and is therefore absorbed in the renormalisation constant. With that, \labelcref{eq:symmetricflow} is consistent with the flowing on-shell renormalisation condition \labelcref{eq:onshellrenSymmetric}.

\subsection{Broken phase} 
\label{sec:BrokenPhase}

In the broken phase with $\phi_0\neq 0$ we use a Taylor expansion about $\phi^2=\phi_0^2$ for the effective potential,
\begin{align}
	V_\textrm{eff}(\phi) =   \sum_{n=2}^\infty \frac{\lambda_{n}}{2 n!} \left( \phi^2-\phi_0^2 \right)^n\,. 
\label{eq:VeffBroken}
\end{align}
At vanishing momentum and constant fields, the correlation functions derived from the effective action $\Gamma[\phi_0]$ coincide with the moments of the effective potential. We consider $n$-point functions for $n\leq 4$ with 
\begin{align}\nonumber
&\Gamma^{(2)}[\phi_0]\left(p=0 \right)= \frac13 \lambda_2 \phi_0^2 \,, \\[1ex]\nonumber
&\Gamma^{(3)}[\phi_0]\left(p=0 \right) =  \,\lambda_{2}\, \phi_0+\frac{\lambda_3}{15}\,\phi_0^3\,, \\[1ex]
&\Gamma^{(4)}[\phi_0]\left(p=0 \right) = \lambda_{2}+ \frac25  \lambda_3 \,\phi_0^2+\frac{1}{105} \lambda_4\, \phi_0^4\,, 
\label{equ:full3and4}
\end{align}
In contrast to the symmetric phase discussed in \Cref{sec:SymPhase}, also higher order terms with couplings $\lambda_n$ contribute due to $\phi_0\neq 0$. For this reason we have indicated the $\phi_0$-dependence of $\Gamma^{(n)}$ in \labelcref{equ:full3and4}. As discussed below \labelcref{eq:prop}, we generically drop the $\phi_0$-dependence for the sake of readability, it is implicitly assumed that all expressions are evaluated at $\phi_0$.  

As a consequence of \labelcref{eq:Approxn>4}, all expansion coefficients $\lambda_n$ with $n \geq 3$ vanish. The three and four-point couplings are then given by 
\begin{align}
	\Gamma^{(3)}(0)=\Gamma^{(4)}(0)\phi_0 \,, \qquad 
	\lambda_2=\Gamma^{(4)}(0) \, .
\label{eq:lambdaphi-4}
\end{align}
With \labelcref{eq:mcurv} we can express the minimum of the effective potential in terms of the curvature mass and $\lambda_2$, yielding
\begin{align}
\phi_0^2= \frac{3 m_\tinytext{curv}^2}{\Gamma^{(4)}(0)}\,,
\label{eq:phi0onshell}
\end{align}
Using \labelcref{eq:phi0onshell}, the three-point function is expressed in terms of the full two- and four-point functions at vanishing momentum,
\begin{align} 
	\Gamma^{(3)}(0) = & \, \sqrt{3 \, \Gamma^{(4)}(0)}\,m_{\tinytext{curv}} \,. 
\label{eq:3vertex}
\end{align} 
Evidently, in the classical limit with $Z_\phi=1$ and $\tilde\rho_k=0$, the curvature mass agrees with the pole mass. This limit is approached for asymptotically large pole masses, where the effective coupling $\lambda_\phi/m_\tinytext{pole}$ tends towards  zero. Hence, the ultraviolet effective potential $V_\tinytext{UV}(\phi)$ with $k=\Lambda\to\infty$ is augmented with a classical dispersion with $\mu=-\Lambda^2$ and the initial (classical) coupling $\lambda_2=\lambda_\phi$, 
\begin{align}
	V_\tinytext{UV}(\phi) = \frac{1}{4!} \lambda_\phi \left( \phi^2-\phi_0^2\right)^2\,, \qquad \phi_0^2 =\frac{6\Lambda^2}{\lambda_\phi} \,, 
\label{eq:VeffUVbroken}
\end{align}
for $\phi^2\geq \phi_0^2$. The initial curvature and pole mass are  given by 
\begin{align}
m_\tinytext{pole}^2 =m_\tinytext{curv}^2 = 2\Lambda^2 \, .
\end{align}
With these approximations, the real-time flow of the two-point function in the broken phase, derived in \Cref{sec:FlowBroken}, reads
\begin{align} \nonumber
	\partial_t\Gamma^{(2)}(\omega_+^2) =& \, \dot{\mathcal{R}} \left( \, D_\tinytext{pol}(\omega_+^2) -\frac12 D^{\tinytext{dyn}}_\tinytext{tad}(\omega_+^2)\right)  \\[1ex] 
	&+  A\, D_\tinytext{fish}(\omega_+^2)+ 4 k^2- \partial_t \hat{S}^{(2)}_{\text{ct}}\,.
\label{eq:CSflowInversePropMink}
\end{align}
The prefactors are given in \labelcref{eq:final2pointflowREN} and $\hat{S}^{(2)}_{\text{ct}}$ is given by 
\begin{align}
\hat{S}^{(2)}_{\text{ct}} = \text{diagramms}\left(p^2=-2k^2\right) \,.
\end{align}
Additionally to the polarisation topology, we note that flow equation in the broken phase differs from that in the symmetric phase. The constant part of \labelcref{eq:CSflowInversePropMink} carries an additional factor of 2. This resembles the additional factor 2 of the squared pole-mass in the broken phase compared to its symmetric phase counterpart. Also the prefactor of the tadpole diagram deviates from the symmetric case, since it includes the implicit $k$-dependence of the internal lines via the flowing physical minimum.

\subsection{Resum\'{e}} \label{eq:Resume}

In both phases, we have a positive curvature mass $m_\tinytext{curv}>0$ on the equation of motion $\phi_0$. Its value is related to the pole mass $m_\tinytext{pole}=k$ in the symmetric, and $m_\tinytext{pole}=2k$ in the broken phase. The difference between the flows is the existence of vertices $\Gamma^{(2n+1)}$ in the broken phase. They are proportional to sums of powers of $\phi_0$, see \labelcref{equ:full3and4}, and hence vanish in the symmetric phase.  Specifically, the flow of the two-point function in the broken phase contains the diagrammatic topology of a vacuum polarisation, see \Cref{Diag:flowingMinimum2Point}. 

This leads us to the following structure: the CS flows are initiated deep in the symmetric and deep in the broken phase for large pole masses and a given classical coupling $\lambda_\phi$, see \labelcref{eq:VeffUVsymmetric,eq:VeffUVbroken} respectively. For the broken phase this entails, that also the field expectation value at the initial scale is large as it scales with $\Lambda$, see \labelcref{eq:phi0onshell,eq:VeffUVbroken}. Then, the pole mass is successively lowered and for $k=0$ one reaches the phase transition point from both sides. In particular, the flows do not leave the broken or symmetric phase. This is in seeming contradiction to the standard fRG picture in a scalar theory,  where flows in the broken phase may end up in the symmetric phase, and those in the symmetric phase end up deeper in the symmetric phase. This apparent contradiction is resolved by the fact, that $\phi_0$ in the standard fRG is defined from the subtracted EoM. There, the trivial cutoff flow, which is $\propto k^2 \phi^2$, is subtracted from the effective potential, and one recovers physics only in the limit $k\rightarrow 0$.

\section{Results}
\label{sec:Results}

\begin{figure*}[ht]
	\centering
	\begin{subfigure}{.47\linewidth}
		\centering
		\includegraphics[width=\textwidth]{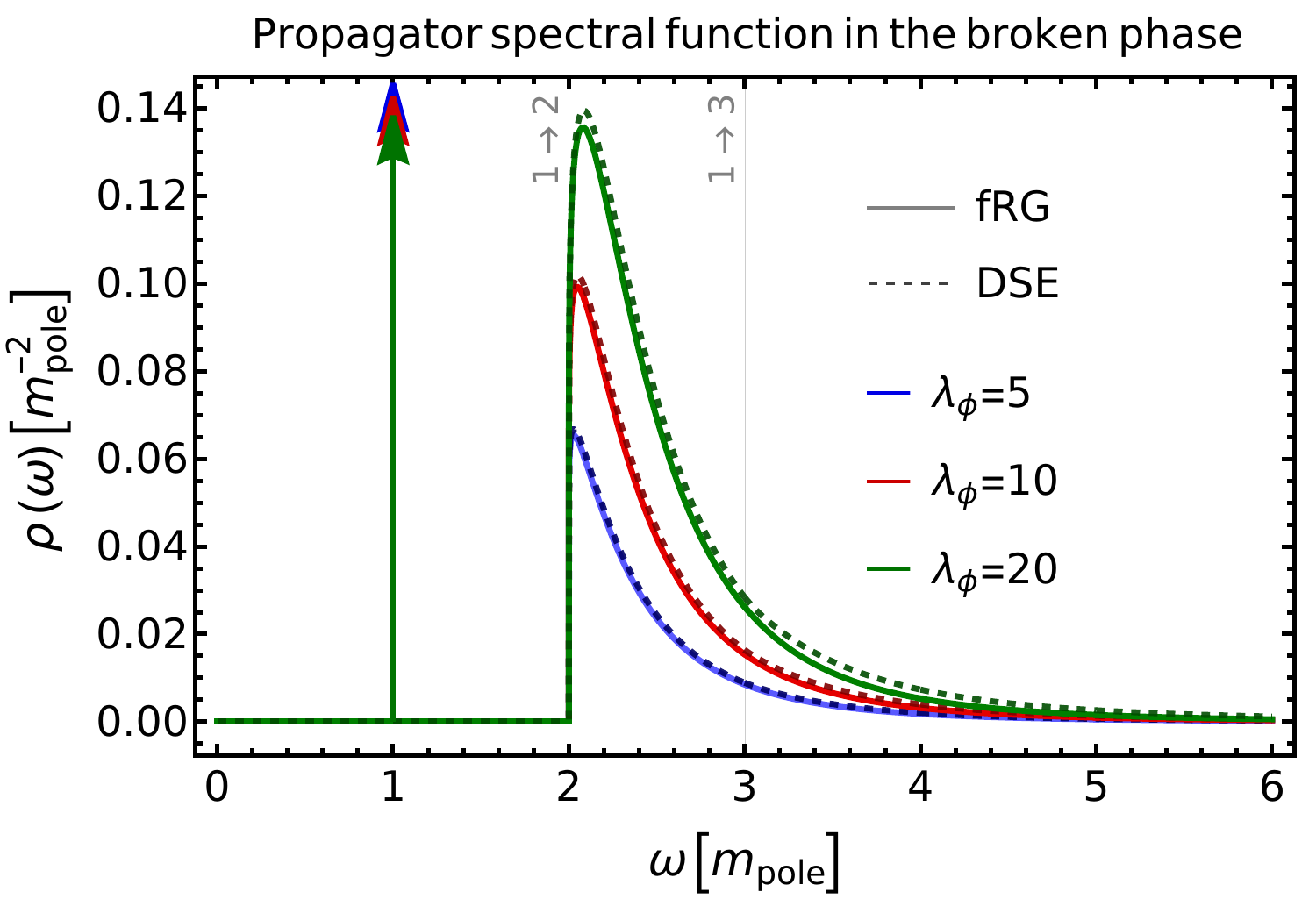}
		\subcaption{Spectral function of the propagator. The $1\rightarrow 2$ and $1\rightarrow3$ particle scattering onsets are indicated in grey.\hspace*{\fill}}
		\label{figPropSpecBroken}
	\end{subfigure}%
	\hspace{0.05\linewidth}%
	\begin{subfigure}{.47\linewidth}
		\centering
		\includegraphics[width=\textwidth]{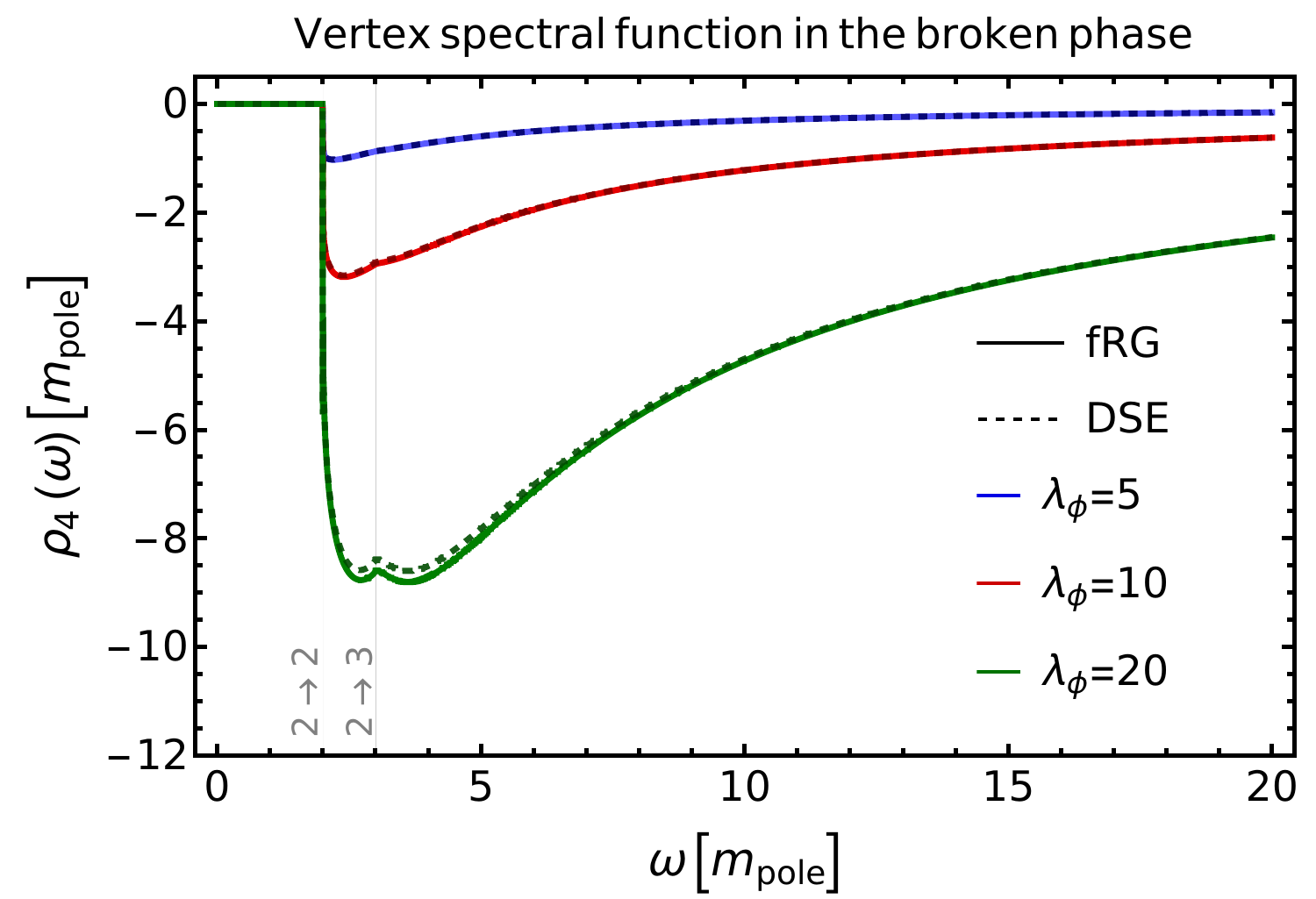}
		\subcaption{Spectral function of the four-point function. The $2\rightarrow 2$ and $2\rightarrow 3$ particle scattering onsets are indicated in grey.\hspace*{\fill}}
		\label{fig:4PSpecbroken}
	\end{subfigure}
	\caption{Spectral functions as a function of Minkowski frequency in comparison to DSE results from \cite{Horak:2020eng}. In contrast to~\Cref{fig:3dplot}, all quantities are measured in units of the pole mass to facilitate the comparison with the DSE results. \hspace*{\fill}}
	\label{fig:Resultsresummedvertices}
\end{figure*}

In this section, we present results for the non-perturbative spectral functions of the scalar propagator in the symmetric and broken phase. The discussion of the numerical implementation is deferred to \Cref{sec:Numerics}. The results allow for an investigation of the scattering processes in both phases. The present results are in remarkable quantitative agreement with that obtained with the spectral DSE in \cite{Horak:2020eng}. This agreement of the spectral functions from these two different functional approaches hold true for a large range of effective couplings $\lambda_\phi/m_\phi$, see \Cref{figPropSpecBroken}. In this coupling regime this agreement  provides a non-trivial reliability check for both functional approaches, thus decreasing the respective systematic error. This error analysis is augmented with a comparison of the present advanced truncation with the classical vertex approximation in \Cref{sec:classicVertices}.

\subsection{Symmetric phase}
\label{sec:Resultssymmetric}
\begin{figure}
	\begin{minipage}{0.47\textwidth}
		\centering 
		\includegraphics[width=\textwidth]{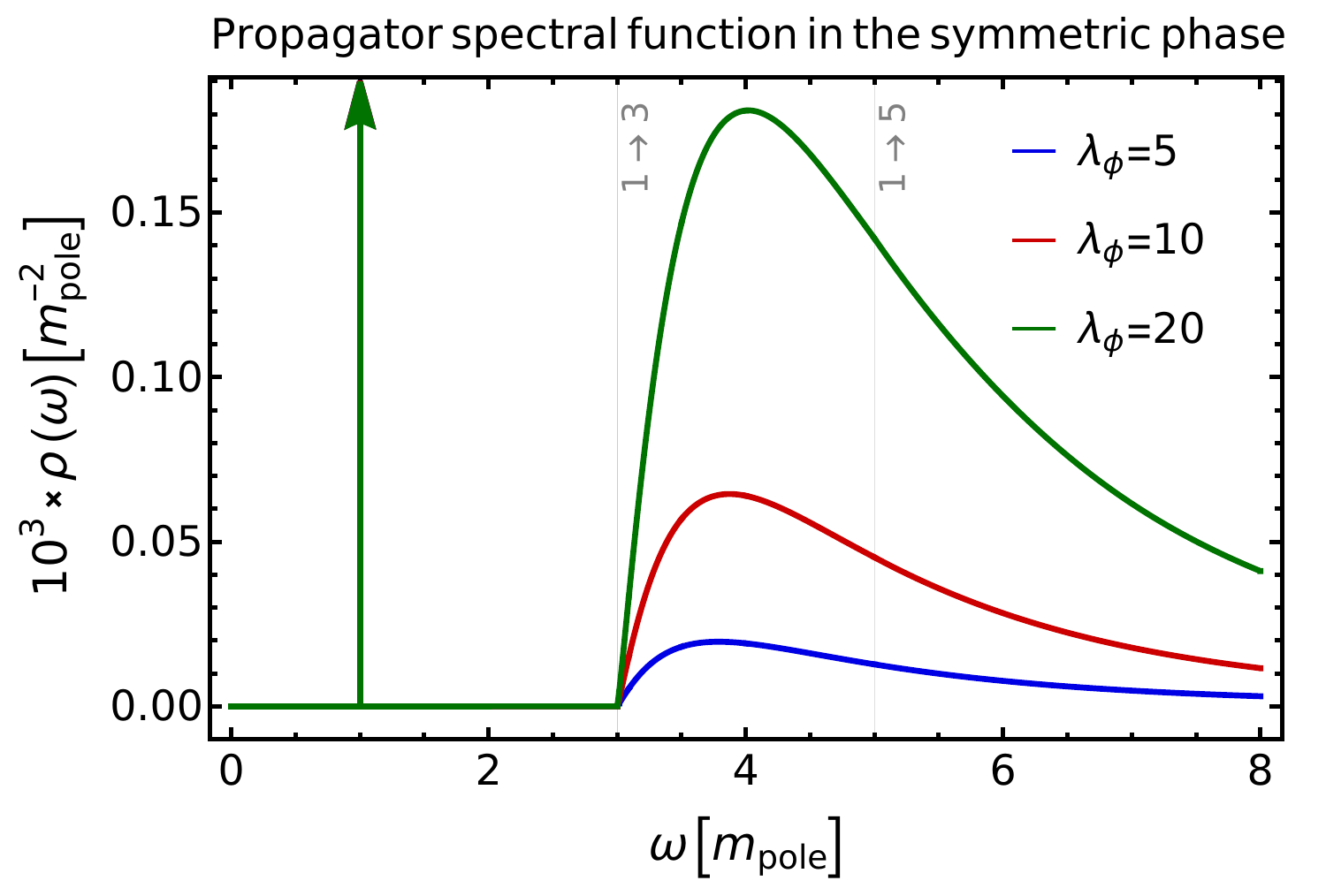}\vspace*{-.7mm}	
		\caption{Spectral functions for vanishing field value as a function of the frequency. All quantities are measured in units of the pole mass. $1\rightarrow 3$ and $1\rightarrow 5$ onsets are indicated in grey. \hspace*{\fill}}
		\label{fig:PropSpecSym}
	\end{minipage}
\end{figure}
\begin{table}[t]
\begin{minipage}{0.44\textwidth}
\centering\vspace*{-.2cm}
\renewcommand{\arraystretch}{1.6}
\begin{tabular}{c|c|c|c}
$ \lambda_\phi /m_\tinytext{pole}$ & $1/Z_\phi$ (fRG) & $1/Z_\phi$ (DSE) & $1/Z_\phi$($\phi_0=0$) \\
\hline
$ 5 $ & $0.971$ & $0.969$ & $0.9998$\\
\hline
$ 10 $ & $0.950$ & $0.945$ &$0.9995$\\
\hline
$20 $ & $0.921$ & $0.907$ &$0.9986$
\end{tabular}
\caption{Amplitudes $1/Z_\phi$ of the pole contribution for given effective couplings, corresponding to the scattering tails displayed in \Cref{figPropSpecBroken} and \Cref{fig:PropSpecSym}.\hspace*{\fill} }
\label{Tab:Residues_broken}
\end{minipage}
\end{table}

%

In the symmetric phase with $\phi_0=0$ we are left with the tadpole diagram in the flow of the two-point function \labelcref{eq:CSflowInverseProp}. The resummation~\labelcref{eq:resummed-four-pt-func} introduces a non-trivial momentum dependence to the four-point function and, in consequence, also to the tadpole diagram. This allows to calculate the propagator spectral function in the symmetric phase, \ie, at vanishing field value, where the polarisation diagram is absent. For the respective flow equation on the real frequency axes see \labelcref{eq:symmetricflow}.
The resulting spectral function is shown in~\Cref{fig:PropSpecSym}. In the symmetric phase, the scattering continuum starts at $3 m_\tinytext{pole}$. As mentioned above, the dynamic tadpole contribution~\labelcref{eq:tadpole} carries the momentum structure of the polarisation diagram, resembling the $s$-channel structure of the four-vertex. Still, the onset of its imaginary part is at thrice the pole mass, since the bubble resummed vertex represents a series of $2\rightarrow 2 $ scatterings which leads to a generic two-particle onset of $\rho_4$. The quantum corrections to the symmetric phase propagator are small compared to the broken phase. The amplitude on the mass pole is close to one compared to the respective values in the broken phase, see \Cref{Tab:Residues_broken}. This is expected, since the first dynamic contribution is of two-loop order and corresponds to the sunset topology.

\subsection{Broken phase}
\label{sec:ResultsBroken}

In the broken phase the condensate is non-vanishing, $\phi_0\neq 0$. To compute the spectral function, the flow equation is evaluated on the real frequency axes see \labelcref{eq:CSflowInversePropMink}.

The $N$-particle onset positions of the spectral scattering tail are governed by the imaginary part of \labelcref{eq:CSflowInversePropMink}. For the polarisation diagrams, where only propagators come into play, the flow exhibits an onset at the sum of the two mass-poles. In contrast, the contribution of the tadpole leads to an onset at thrice the pole-mass, as the four-point spectral function only consists of a scattering continuum starting at $2 m_\tinytext{pole}$, cf.~\Cref{fig:4PSpecbroken}. A more detailed discussion of the scattering onsets in general is found in \Cref{sec:analyticintegrands} and specifically for the tadpole in \Cref{sec:Resultssymmetric}. 

In ~\Cref{figPropSpecBroken} the spectral function from the current fRG approach is compared to spectral DSE results from~\cite{Horak:2020eng}. Every quantity is measured relative to the respective pole mass to facilitate comparison with the DSE results. This allows to compare the relative magnitude of the scattering continua for different coupling strengths. For effective couplings $\lambda_\phi/m_\tinytext{pole}\lesssim 20$, the spectral weight of the scattering continuum is sub-leading, as can be inferred from the combination of~\Cref{fig:3dplot_Z} and the sum rule~\labelcref{eq:specnormForAllk}. The amplitudes of the pole contributions are listed in \Cref{Tab:Residues_broken}.

We find a remarkable agreement of both methods in the tested coupling range. For effective couplings $\lambda_\phi/m_\tinytext{pole}\approx 20$, the deviations start growing, specifically at the thresholds. Deviations between both methods arise due to differences in the resummation structure of the two functional equations in the current truncation. The convergence of functional techniques for a large range of couplings is non-trivial and strengthens our confidence in spectral functional approaches.

In general, the tail of the propagator spectral function is enhanced for stronger couplings, while the residue of the mass pole decreases as the scattering states become more accessible due to the rising dimensionless interaction strength. The three- and higher $N$-particle onsets are graphically not visible in the full spectral functions of~\Cref{figPropSpecBroken}, but present in the data. In the limit of large couplings we expect the three-particle onset to become more pronounced as the tadpole contribution becomes large.

The four-point spectral function shown in~\Cref{fig:4PSpecbroken} consists of only a negative scattering tail corresponding to a $ 2\rightarrow2$ scattering process. For higher couplings, the three-particle onset becomes visible. The different suppression of higher $N$-particle thresholds in the propagator- and four-point spectrum are explained by dimensional analysis. While for the propagator spectral function, higher $N$-particle onsets are suppressed by their squared energy threshold, the four-point spectral function decays only with $\lambda^{-1}$, leading to a suppression linear in their respective energy thresholds. In both cases, four-particle or higher onsets are strongly suppressed, since they come with at least one additional loop each.

\begin{figure}
	\begin{minipage}{0.45\textwidth}
		\centering
		\includegraphics[width=\textwidth]{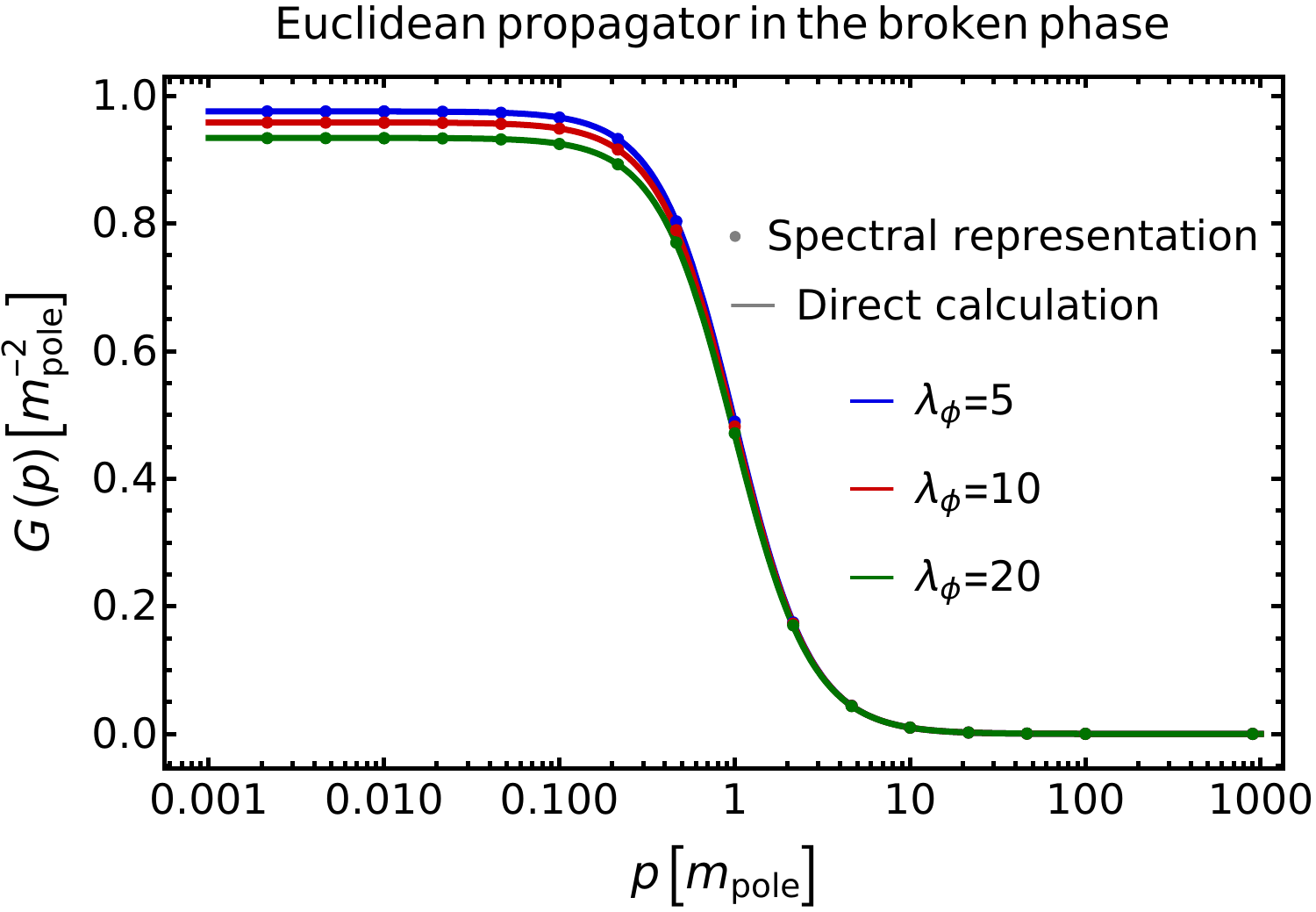} 
		\caption{Propagator as function of Euclidean frequency. This result serves as a cross-check between a direct computation via the flow and a calculation using the spectral functions.\hspace*{\fill}}
		\label{fig:ResultsresummedverticeseuclLEFT}
	\end{minipage}\hspace*{0.05\linewidth}
\end{figure}

\Cref{fig:ResultsresummedverticeseuclLEFT} shows the Euclidean propagators corresponding to the spectral functions of~\Cref{figPropSpecBroken}. As a cross-check, we compare the Euclidean propagator calculated from the spectral representation to the propagator directly obtained from the integrated Euclidean flow. We find the spectral representation to hold.
%

\section{Conclusion}\label{sec:Conclusions}

In the present work, we computed single particle spectral functions of a scalar $\phi^4$-theory within the spectral functional renormalisation group approach, put forward in \cite{Braun:2022mgx}. This approach leads to renormalised spectral flows with flowing renormalisation, and facilitates a fully self-consistent computation of non-perturbative spectral functions.
We derived full flow equations for the inverse propagator in both, the symmetric and broken regime of the theory, for a detailed discussion see \Cref{sec:flowingMinimum,eq:Resume}.
 
Our setup is manifestly Lorentz invariant and sustains the causal properties of the theory throughout the flow. Every point on the Callan-Symanzik RG trajectory is a physical theory of scale $k$. Trajectories in the symmetric and broken regime each start from an infinitely heavy theory in the respective phase and meet at the phase boundary in the strongly interacting massless limit of the theory, see \Cref{sec:funCS} and \Cref{fig:FlowLeftRight}. Thereby, our setup avoids flows through the strongly interacting phase transition regime, which are usually present in momentum cutoff flows. This minimises the systematic error stemming from the strong dynamics in the vicinity of a phase transition, where the flows are highly sensitive to truncation artefacts. Furthermore, the implementation of a flowing renormalisation condition eliminates the need of fine-tuned initial conditions and allows for monotonous mass flows.

The explicit results in the broken phase are in impressive agreement with those obtained in \cite{Horak:2020eng} within the spectral DSE, see \Cref{sec:ResultsBroken}. This affirms the reliability of the spectral functional approach for the non-perturbative computation of fundamental Minkowski spacetime correlation functions.
 
In contrast to DSE, the fRG approach captures average momentum dependencies of vertices via their scale dependence. This allows to include non-trivial vertex dynamics without resorting to intricate spectral representations of higher correlation functions. Furthermore, the current spectral fRG approach is straightforwardly and easily extended to include the flow of the full effective potential. This work represents an important step towards unravelling real-time correlations in QCD from first principles with spectral functional approaches. We hope to report on respective results in the near future.

\section{Acknowledgements}

We thank Gernot Eichmann, Lorenz von Smekal, and Johannes Roth for discussions and collaborations on related projects. This work is done within the fQCD-collaboration~\cite{fQCD}, and we thank the members for discussion and collaborations on related projects. This work is funded by the Deutsche Forschungsgemeinschaft (DFG, German Research Foundation) under Germany's Excellence Strategy EXC 2181/1 - 390900948 (the Heidelberg STRUCTURES Excellence Cluster) and under the Collaborative Research Centre SFB 1225 (ISOQUANT). JH and FI acknowledge support by the Studienstiftung des deutschen Volkes. NW~is supported by the Hessian collaborative research cluster ELEMENTS and by the DFG Collaborative Research Centre "CRC-TR 211 (Strong-interaction matter under extreme conditions)".

\appendix

\begingroup
\allowdisplaybreaks

\section{Spectral diagrams}\label{sec:specDiags}

The general spectral form of the diagrams in \Cref{fig:2pointflow} and \Cref{Diag:threepoint} is given by 
\begin{align}  \nonumber
	&D_\tinytext{tad}(p^2) =   \left(\prod^2_{i=1}\int_{\lambda_i}\rho(\lambda_i)\right)\, \mathcal{L}_\tinytext{tad}(\vec{\lambda},p^2) \,,	\\[1ex]  \nonumber
	&D_\tinytext{pol}(p^2) = \left(\prod^3_{i=1}\int_{\lambda_i}\rho(\lambda_i)\right)\mathcal{L}_\tinytext{pol}(\vec{\lambda},p^2)	 \,,	\\[1ex]  
	&D_\tinytext{fish}(p^2) =   \left(\prod^2_{i=1}\int_{\lambda_i}\rho(\lambda_i)\right)\, \mathcal{L}_\tinytext{fish}(\vec{\lambda},p^2)\,,
\label{eq:generalDiags}
\end{align}
with $\vec{\lambda}=(\lambda_1,..,\lambda_n)$ denoting the ordered vector of spectral masses. The loop structure reads
\begin{align}\nonumber
		&\mathcal{L}_\tinytext{tad}(\vec{\lambda},p^2)= \int_q \frac{\Gamma^{(4)}[p,q,-p]}{(q^2+\lambda_1^2)(q^2+\lambda_2^2)}\,, \\[1ex]\nonumber
		&\mathcal{L}_\tinytext{pol}(\vec{\lambda},p^2)=  \int_q \frac{\left(\Gamma^{(3)}[p,q]\right)^2}{(q^2+\lambda_1^2)(q^2+\lambda_2^2)\left((p-q)^2+\lambda_3^2\right)} \,,\\[1ex] 
&\mathcal{L}_\tinytext{fish}(\vec{\lambda},p^2)=  \int_q \frac{\Gamma^{(3)}[p,q]\,\Gamma^{(4)}[p,q,0]}{\left(q^2+\lambda_1^2\right)\left((p-q)^2+\lambda_2^2\right)} \,,
\end{align}
where the vertex functions are not specified yet, and we have dropped the field argument for readability. With the approximations discussed in \Cref{sec:phi4}, we have fixed all correlation functions and we can compute the final expressions for the diagrams. For the fish diagram, the four-point function is connected to the constant scale derivative of the field. With~\labelcref{eq:3vertex} and~\labelcref{eq:lambdaphi-4}, the polarisation- and fish diagram of the flow equation in the broken phase read
\begin{subequations}\label{eq:Dpol}
\begin{align}\nonumber
	D_\tinytext{pol}[p^2]=& \left(\Gamma^{(3)}\right)^2 \hspace{-1.3mm} \int_{\vec\lambda} \,\rho(\lambda_1)\rho(\lambda_2)\rho(\lambda_3)\, I_\tinytext{pol}(\vec{\lambda},p^2) \,, \\[2ex]
	D_\tinytext{fish}[p^2]=& \frac{\left(\Gamma^{(3)}\right)^2}{\phi_0} \hspace{-1.3mm} \int_{\vec\lambda} \,\rho(\lambda_1)\rho(\lambda_2)\, \tilde{I}_\tinytext{pol}(\vec{\lambda},p^2) \,,
 \label{eq:ConstVertexpoldiag}
\end{align}
where
\begin{align}\label{eq:Ipol} \nonumber
	I_\tinytext{pol}(\vec{\lambda},p^2)=& \int_q \frac{1}{(q^2+\lambda_1^2)(q^2+\lambda_2^2)((q-p)^2+\lambda_3^2)} \,,\\[2ex]
	\tilde{I}_\tinytext{pol}(\vec{\lambda},p^2)=& \int_q \frac{1}{(q^2+\lambda_1^2)((q-p)^2+\lambda_2^2)} \,,
\end{align}
\end{subequations}
and $\vec{\lambda}$ is the ordered vector of spectral parameters, one for each internal propagator in the diagram. We denoted the loop structure of the fish diagram as $\tilde{I}_\tinytext{pol}$, since it carries the momentum structure of a DSE polarisation diagram. The loop integrals $I$ and $\tilde{I}$ are evaluated analytically and given in terms of real and imaginary frequencies in~\Cref{sec:analyticintegrands}. 

Substituting the four-vertex in~\labelcref{eq:CSflowInverseProp} with the respective spectral representation~\labelcref{eq:specrep4point}, the constant classical part of the tadpole is absorbed by renormalisation. The remaining dynamical part of the tadpole diagram reads 
\begin{align} 
	D^{\tinytext{dyn}}_\tinytext{tad}(p) =&   \int_{\vec\lambda} \, \rho(\lambda_1)\rho(\lambda_2)\rho_{4}(\lambda_3)\, I_\tinytext{pol}(\vec{\lambda},p^2) \,. 
\label{eq:tadpole}
\end{align} 
The four-point spectral function $\rho_4$  \labelcref{eq:tadpole} is obtained from \labelcref{eq:resummed-four-pt-func} in both phases. As discussed below \labelcref{eq:resummed-four-pt-func}, there are further diagrams with two or four three-point functions contributing to $\rho_4$. These diagrams are suppressed by roughly an order of magnitude.

\section{Renormalised flow of the two-point function on the physical minimum}\label{sec:FlowBroken}

In this appendix, we provide details on the derivation of the flow equation in the broken phase~\labelcref{eq:FullFlowGammaFinal1}. In particular, we explain the crucial role of the three-point function in~\labelcref{eq:FullFlowGamma2} for obtaining one-loop perturbation theory as leading order behaviour. We show that the flow of the two-point function has the expected sign, see~\labelcref{eq:treelevelflow}, if we include the flow of the minimum correctly and that the on-shell renormalisation condition~\labelcref{eq:onshellrenBroken} can be imposed consistently.

The flow equation in the broken phase reads 
\begin{subequations}
	\begin{align}\nonumber 
		&\hspace*{-.14cm}\frac{d}{dt}\Gamma^{(2)}[\phi_0](p)=\;\left(\partial_t\phi_0\right)S^{(3)}[\phi_0] -\left(2-\eta_\phi\right)Z_\phi k^2\\*[2ex]
		& \hspace{.5 cm} +\dot{\mathcal{R}}  \left[-\frac12 D_\tinytext{tad} + D_\tinytext{pol}\right]  -\partial_t \phi_0 D_{\tinytext{fish}}-\partial_t S_\tinytext{ct}[\phi_0] \, ,
		\label{eq:FullFlowGammaFinal1appendix}
	\end{align}
	where
	\begin{align} \label{eq:rdot-appendix}
	\dot{\mathcal{R}}=\left(\partial_t \phi_0 \Gamma^{(3)}[\phi_0] - \left(2-\eta_\phi \right) Z_\phi k^2 \right) \,,
	\end{align}
\end{subequations}
as derived in \Cref{sec:flowingMinimum}.

To renormalise the flow of the two-point function and show how the correct sign of the flow is recovered by the inclusion of $\partial_t \phi_0$, we first separate the tree-level and loop-induced running of the prefactors of the diagrams in~\labelcref{eq:FullFlowGammaFinal1appendix}. To that end, we start with the full three-point function in~\labelcref{eq:rdot-appendix}. The separation into trivial and loop-induced RG running can be made apparent by introducing a corresponding split of the curvature mass~\labelcref{eq:mcurv} via $\Delta m_\tinytext{curv}^2 =  m_\tinytext{curv}^2 -  2 Z_\phi k^2$
\begin{subequations}\label{eq:dtphi0threepoints}
\begin{align}\label{eq:dtphi0ConstGamma3}
&\partial_t \phi_0 \Gamma^{(3)} = \frac12 \left(\partial_t \phi_0^2\right) \Gamma^{(4)} \\[2ex] \nonumber
&= \frac32 \left( \partial_t\Delta m_\tinytext{curv}^2 + 2Z_\phi(2-\eta_\phi)k^2-  m_\tinytext{curv}^2 \, \frac{\partial_t\Gamma^{(4)}}{\Gamma^{(4)}}\right) \,.
\end{align}
In the first line, we related the three- and four-point function by~\labelcref{eq:lambdaphi-4}, and we used \labelcref{eq:phi0onshell} in the second step. 

For the classical three-vertex in~\labelcref{eq:FullFlowGammaFinal1appendix}, we analogously obtain
\begin{align}\label{eq:dtphi0S3}
 \partial_t \phi_0 S^{(3)} = \frac{\lambda_\phi}{\Gamma^{(4)}} \left(\partial_t \phi_0 \Gamma^{(3)} \right) \,.
\end{align}
\end{subequations}
With~\labelcref{eq:ConstVertexpoldiag}, the fish diagram in~\labelcref{eq:FullFlowGammaFinal1appendix} carries a prefactor proportional to 
\begin{align} \label{eq:partialphi/phi}
	\frac{\partial_t \phi_0}{\phi_0}&= \frac{\partial_t m_\tinytext{curv}^2}{m^2_\tinytext{curv}}- \frac{\partial_t \Gamma^{(4)}}{\Gamma^{(4)}}\,.
\end{align}
Inserting \labelcref{eq:dtphi0threepoints,eq:partialphi/phi} in \labelcref{eq:FullFlowGammaFinal1appendix}, we can write down the final flow equation for the two-point function in the broken phase,
\begin{subequations}
\begin{align}
\nonumber 
	\frac{d}{dt}\Gamma^{(2)}[\phi_0](p) =&\; \dot{\mathcal{R}} \left( - \frac12 D_\tinytext{tad} + D_\tinytext{pol} \right) + A\, D_\tinytext{fish}  \\[2ex] 
&\hspace*{-1cm}+	 B\, (2-\eta_\phi )Z_\phi k^2 + C - \partial_t S_\tinytext{ct}^{(2)} \,,
\label{eq:final2pointflow}
\end{align}
with 
\begin{align}\nonumber
\dot{\mathcal{R}}&= 2 Z_\phi\left(2-\eta_\phi\right)\, k^2 + \frac32 m_\tinytext{curv}^2 \left(\frac {\partial_t \Delta   m_\tinytext{curv}^2}{m_\tinytext{curv}^2} -	     \frac{\partial_t\Gamma^{(4)}}{\Gamma^{(4)}}\right)\,,\\[2ex]\nonumber
A&=-\frac{\phi_0}{2} \left(\frac{\partial_t m_\tinytext{curv}^2}{m^2_\tinytext{curv}}- \frac{\partial_t \Gamma^{(4)}}{\Gamma^{(4)}} \right) \,,\\[2ex]\nonumber
B&=\left(\frac{3 \lambda_\phi}{\Gamma^{(4)}} -1\right)\,,\\[2ex]
C&=\frac32 m_\tinytext{curv}^2 \frac{\lambda_\phi}{\Gamma^{(4)}} \left(\frac{\partial_t{\Delta m_\tinytext{curv}^2}}{m_\tinytext{curv}^2} - \frac{\partial_t \Gamma^{(4)}}{\Gamma^{(4)}} \right)\,.
\label{eq:ABCD} 
\end{align}
\end{subequations}
The diagrams $D_\tinytext{pol},\,D_\tinytext{fish}$ and $D_\tinytext{tad}^{\tinytext{dyn}}$ are given in \labelcref{eq:Dpol,eq:tadpole}, respectively. 

The prefactor of the first term in~\labelcref{eq:ABCD}, $\dot{\mathcal{R}}$, carries the scale dependence of the internal propagators on the physical minimum. It takes the role of the regulator derivative in usual fRG applications. The first term of $\dot{\mathcal{R}}$ and the third term of the flow~\labelcref{eq:final2pointflow} are explicitly proportional to $k^2$, and have the same structure as the respective contributions in~\labelcref{eq:CSflowInverseProp}, where the flow equation is evaluated at arbitrary values of the field.
However, it carries an additional relative factor $-2$, stemming from the three-point function, which is proportional to $\partial_t\phi_0 \Gamma^{(3)}$. In total, the tree-level term of $\dot{\mathcal{R}}$ is positive and can be written as the $t$-derivative of $2Z_\phi k^2$. This reflects the positivity of the physical pole mass. At one-loop order, it reduces to $4k^2$, resembling the $k$-dependence of a classical propagator with $m_\phi=2k^2$, see~\labelcref{eq:treelevelflow}. 

The same holds true for the constant part of the flow, given by the third and fourth term of~\labelcref{eq:final2pointflow}. Complementary to $\dot{\mathcal{R}}$, these encode the explicit running of the mass. At tree-level, this running reduces to $4k^2$, which is nothing but the flow of the classical (squared) mass on the physical minimum. Hence, the deviation of the constant part of the flow from the classical running is, as expected, of one-loop order and beyond and can be absorbed in the renormalisation constant. With that, we can consistently impose our renormalisation condition.

The second term of \labelcref{eq:final2pointflow} is proportional to the fish diagram. Note that $D_\tinytext{fish}$ carries a factor $1/\phi_0$, which cancels the respective factor in $A$.  At one-loop order, it carries the running of the classical three-point function. Together, $D_\tinytext{fish}$ and $D_\tinytext{pol}$ can be written as the total derivative of the (one-loop) vacuum polarisation. 

We now apply our renormalisation condition~\labelcref{eq:onshellrenBroken} to the flow~\labelcref{eq:final2pointflow}, for which it translates into the condition 
\begin{align} \label{eq:renorm-cond_flow}
	\partial_t m_\tinytext{pole}^2= 4k^2 \,.
\end{align}
This specifies our counterterm flow and leads us to the final renormalised flow equation in the broken phase:
\begin{align}
	\nonumber 
	\frac{d}{dt}\,\Gamma^{(2)}(p) =& \,\dot{\mathcal{R}}\left(  \, D_\tinytext{pol}(p) - \frac12 D^{\tinytext{dyn}}_\tinytext{tad}(p)\right)  \\[1ex] 
		&+ A\, D_\tinytext{fish}(p)+ 4 k^2- \partial_t \hat{S}^{(2)}_{\text{ct}}\,,
	\label{eq:final2pointflowREN}
\end{align}
where we have split the tadpole in a constant and dynamical $p$-dependent part defined by~\labelcref{eq:tadpole}, via the spectral representation~\labelcref{eq:specrep4point} of the four-vertex.
Furthermore, we have dropped all constants in $p$ of order $(\lambda_\phi k)$, including the constant part of the tadpole, as they are subtracted by the renormalisation constant implicitly specified by~\labelcref{eq:renorm-cond_flow}. The remaining $\partial_t \hat{S}_\tinytext{ct}^{(2)}[\phi_0]$ now comprises only the counterterms of the diagrammatic contributions, where the renormalisation scale is the pole mass.

\section{Flow of \texorpdfstring{$\phi_0$}{phi0} and critical exponents}\label{sec:Evolutionphi0}

In this appendix, we discuss the evolution of the solution of the EoM, $\phi_0$, in the broken phase. It is given by \labelcref{eq:phi0onshell}. This exact relation depends on $\lambda_2$, which we have identified with $\Gamma^{(4)}$, dropping higher order terms proportional to $\lambda_{3}, \lambda_4$ and $\phi_0$ itself. Implicitly, these terms can be included by solving the flow of $\phi_0$. It is derived from the EoM for constant fields, which is solved for a $k$-dependent $\phi_0$. Acting with a total $t$-derivative on the EoM \labelcref{eq:EoM} leads us to 
\begin{align}\label{eq:dotphi0}
	\partial_t\phi_0=-\frac{\partial_t V_\textrm{eff}^{(1)}(\phi_0)}{V_\textrm{eff}^{(2)}(\phi_0)}=-\frac{\partial_t V_\textrm{eff}^{(1)}(\phi_0)}{m_\tinytext{curv}^2}\,.
\end{align}
The denominator is nothing but the curvature mass squared, while the numerator is given by the first field derivative of the CS equation \labelcref{eq:CSflow-gamma}, evaluated at $\phi_0$. At each flow step, the latter generates higher order terms beyond the approximation \labelcref{eq:Approxn>4}. In summary, if using the flow equation in \labelcref{eq:dotphi0}, we implicitly take into account terms dropped in \labelcref{eq:phi0onshell}. 
In the present approximation the numerator of \labelcref{eq:dotphi0} reads
\begin{align}\label{eq:dotdV}\nonumber
\partial_t{V}^{(1)}[\phi_0] &=  \phi_0 \, \left(2-\eta_\phi\right) Z_\phi k^2\left[\frac12 D_\tinytext{tad}(0) - 1\right] \\*[1ex] 
 &\hspace{3cm}- \phi_0 \left(\partial_t\Delta m^2\right)\, .
\end{align}
The last term stems from the flow of the counterterm action 
\begin{align}
\partial_t S_\tinytext{ct}[\phi]= \text{Tr} \; \frac12 \left(\partial_t\Delta m^2\right)\phi^2 \, .
\end{align}
Collecting the terms proportional to $\phi_0$ and $\partial_t{\phi_0} $, we arrive at 
\begin{align}\label{eq:flowLogphi}
\partial_t \mathrm{log}(\phi_0)=& \frac{\left(2-\eta_\phi\right) Z_\phi k^2}{m_\tinytext{curv}^2}\, \left( 1 + \mathcal{T}\right)\, ,
\end{align}
where $\mathcal{T}$ comprises the corrections from the tadpole diagram and the counterterm
\begin{align}
\mathcal{T} = -\frac12 D_\tinytext{tad}(0)+  \frac{\partial_t \Delta m^2 }{\left(2-\eta_\phi\right) Z_\phi k^2 }\, .
\label{eq:CalT}
\end{align}
\Cref{eq:flowLogphi} is easily integrated, leading to  
\begin{align}\label{eq:phi0}
\phi_{0}&=\phi_{0,\Lambda}\,\mathrm{exp}\left\lbrace   \int^k_{\Lambda} \frac{dk}{k} \frac{\left(2-\eta_\phi\right) Z_\phi k^2}{m_\tinytext{curv}^2}\, \left( 1 + \mathcal{T}  \right) \right\rbrace \, ,
\end{align}
where $\phi_{0,\Lambda}$ is the classical ultraviolet value of the condensate in the initial UV effective potential, \labelcref{eq:VeffUVbroken}, at the initial large mass $\Lambda$. For smaller pole masses the condensate gets progressively smaller and vanishes for $k=0$.
We can simplify~\labelcref{eq:phi0} further by noting that the squared curvature mass, see \labelcref{eq:phi0onshell}, is conveniently written in terms of the spectral representation 
\begin{align}\label{eq:spec_mcurv}
m_\tinytext{curv}^2 &= \frac{2 Z_\phi k^2}{1+ \int_2^\infty \frac{d\lambda}{\lambda}  \bar \rho(\lambda)} \,,
\end{align}
where we defined the RG-invariant spectral function as
\begin{align}\label{eq:RGrho}
\bar{\rho}\left(\lambda\right) &= \frac{2Z_\phi k^2}{\pi} \rho\left( \sqrt{2k^2}\, \lambda\right)\,.
\end{align}
With \labelcref{eq:spec_mcurv}, the exponent of \labelcref{eq:phi0} reads
\begin{align}\label{eq:phi0exponent}
 \int^k_{\Lambda}\frac{dk}{k} \left[\left(1-\frac{\eta_\phi}{2}\right)\left(1+  \int_2^\infty \frac{d\lambda}{\lambda}  \bar \rho(\lambda)\right)\left(1+\mathcal{T}\right)\right] \, .
\end{align}
For large k, \ie $\lambda_\phi\ll k$ we can approximate $m^2_\text{curv} \approx 2 k^2 \, ,$ and $ \lambda_2 = \Gamma^{(4)}(0) \approx \lambda_\phi$.  The flow of the renormalisation constants is dominated by the tadpole contribution for large cutoff scales. It is of mass dimension 2 and at leading order in $k$ it is proportional to $(\lambda_\phi  k)$. Consequently,  $\frac{\partial_t{\Delta m^2}}{k^2}  \approx 0 $ as well as the tadpole contribution,
\begin{align}\label{eq:Stadfree}
D_\text{tad}= \lambda_\phi &\int \frac{d^3q}{(2\pi)^3} \frac{1}{(q^2 + 2k^2)^2} = \frac{\lambda_\phi}{\sqrt{2}k\, 8 \pi} \,.
\end{align}
 The propagator is then well approximated by the free one, $Z_\phi=1$ and $\tilde\rho=0$ and \labelcref{eq:phi0exponent} reduces to $1+\mathcal{T}\approx 1+ \bar{\lambda}_\phi c$, with the effective coupling $\bar{\lambda}_\phi=\lambda_\phi/k$ and a dimensionless constant $c$. In the limit of large masses, \labelcref{eq:phi0Largek} flows to the classical solution as expected:
\begin{align}\label{eq:phi0Largek}\nonumber
\phi_{0,k} =& \phi_{0,\Lambda}\mathrm{exp}\left\lbrace  \int^k_\Lambda \frac{dk}{k} +   c\,\int^k_\Lambda \frac{dk}{k}\bar{\lambda} \right\rbrace \\
&\longrightarrow\phi_{0,\Lambda}\left(\frac{k}{\Lambda}\right) \, .
\end{align}
Similar equations could be formulated in terms of $\rho=\phi^2$, reflecting the symmetry of the theory.
In O(N)-theories and in the real scalar case such a representation is typically used, as derivatives in $\rho$ project directly on $\lambda_n$ in both phases. While all different formulations are equivalent if the full effective potential is used, they deal differently with the approximation \labelcref{eq:Approxn>4}.

\subsection{Phase transition and critical scaling}\label{sec:CriticalScaling}

Here we provide a qualitative discussion on the scaling limit and use the integrated flow of the physical minimum, $\phi_0$, and the (hyper-)scaling relation~\labelcref{eq:Magentisation} to derive explicit equations for the scaling exponent $\eta_\phi$. The phase transition between the symmetric and broken phase is reached with $k\to 0$ in both phases. In the limit of a vanishing pole mass, $m^2_\tinytext{pole}=2k^2=2|\mu|\to 0$ we are interested in the running of the 'magnetisation' $\bar\phi_0 = \phi_0/\sqrt{Z_\phi}$. The division by $1/\sqrt{Z_\phi}$ eliminates the RG-scaling of the expectation value and leads to the physical observable. In the scaling limit, the magnetisation acquires a scaling form
\begin{align}\label{eq:Magentisation}
	\bar \phi_0\propto \tau^\beta\,,\qquad \beta= \frac12 \nu\left(1 +\eta_\phi\right)\approx 0.3264\,, 
\end{align}
where $\nu \approx 0.6300$ and $\eta_\phi\approx  0.03630$ are the scaling exponents of the three dimensional Ising universality class. The tuning parameter $\tau$ is, in contrast to usual critical theory, not directly proportional to the mass-parameter $\mu$ or $k^2$ as consequence of the flowing on-shell renormalisation. To see that, we consider the scaling form of the correlation length 
\begin{align}\label{eq:xiscaling}
\xi  \propto \tau^{-\nu} \,,
\end{align}
with the mean-field scaling $\nu=\frac12$. In general, the correlation length is inverse proportional to the lowest lying pole of the propagator. Beyond the mean field theory, the correlation length acquires an anomalous scaling in dimensions below $d=4$. With the on-shell renormalisation procedure, this anomalous scaling is hidden and we have
\begin{align}
\xi \propto k^{-1} \,,
\end{align}
for all cutoff scales. This entails that the tuning parameter $\tau$ is related to the pole-mass $m_\tinytext{pole} \propto k$ as 
\begin{align}\label{eq:deftau}
\tau  \propto k^\frac{1}{\nu} \,.
\end{align}
In every flow step, the diagrams of the flow introduce an anomalous scaling to the pole mass, which is subtracted by the counter-term and expresses the renormalisation of the full scaling of the pole mass to the classical one. Hence, the scaling exponent $\nu$ is encoded in the flow of the counter-term in the scaling limit. With~\labelcref{eq:deftau}, the magnetisation~\labelcref{eq:Magentisation} can be rewritten as 
\begin{align}\label{eq:phi0barscaling}
\bar \phi_0\propto k^{\frac{\beta}{\nu}}\,, \qquad \frac{\beta}{\nu} = \frac12 \bigl(1 + \eta_\phi)\,.
\end{align}
The $k$-scaling of the magnetisation is encoded in the \linebreak \texorpdfstring{$k \to 0 $}{kto0} limit of~\labelcref{eq:phi0}. Resolving the brackets, we notice that the first two terms of the exponent~\labelcref{eq:phi0exponent} can be integrated immediately, leading to 
\begin{subequations}\label{eq:phi0_full}
\begin{align}\label{eq:phi0_final}
\phi_{0}&=\phi_{0,\Lambda}\; \sqrt{Z_\phi} \left(\frac{k}{\Lambda}\right)\; \mathrm{exp} \left\lbrace   \int^k_{\Lambda} \frac{dk'}{k'}  \mathcal{D}(k')\right\rbrace \, ,
\end{align}
where we used the definition of $\eta_\phi$. The residual integrand is abbreviated as
\begin{align}\label{eq:calD} \nonumber
\mathcal{D}(k)&= \left(1-\frac{\eta_\phi}{2}\right)\left[\left(1+\mathcal{T}\right) \int_2^\infty \frac{d\lambda}{\lambda}  \bar \rho(\lambda) \right.\\[2ex] 
&\hspace*{2cm}\left. +\mathcal{T} \left(1+\int_2^\infty \frac{d\lambda}{\lambda}  \bar \rho(\lambda)\right)\right] \, ,
\end{align}
\end{subequations}
with $\cal T$ given in \labelcref{eq:CalT}. In the scaling regime we have $k\to 0$ and 
\begin{align}
\bar \phi_0\propto \lim_{k\to 0} k  \,  \mathrm{exp} \left\lbrace   \int^k_{\Lambda} \frac{dk'}{k'}  \mathcal{D}(k')\right\rbrace \, .
\label{eq:phi0bar}
\end{align}
In the limit $k\to 0$ the integral in \labelcref{eq:phi0bar} diverges logarithmically with the prefactor $\mathcal{D}_0=\mathcal{D}(0)$. Then we can read of $\beta/\nu$ and solve~\labelcref{eq:phi0barscaling} for $\eta_\phi$,
\begin{align} 
	\eta_\phi= 1+ 2\mathcal{D}_0\,.
	\label{eq:etaphiD0}
\end{align}
The prefactor $\mathcal{D}_0$ is either computed for $k\to 0$ or is extrapolated when the scaling regime is reached. Alternatively, $\eta_\phi$ can be computed directly from the flow in the scaling limit via it's definition, see \labelcref{eq:eta}. 
The size of the scaling regime can be estimated from the running of the four-point function: For large values of the dimensionless coupling $\lambda_\phi/m_\tinytext{pole} >> 1 $, the loop correction in the denominator of~\labelcref{eq:resummed-four-pt-func} outgrows the constant part,
\begin{align}
\Gamma^{(4)}(p)&=\frac{\lambda_{\phi}}{1+\frac{\lambda_{\phi}}{2} \int_q G(p+q)G(q)} \to \frac{2}{\int_q G(p+q)G(q)} \, ,
\end{align}
and the flow becomes independent of the coupling. Our computational set-up did not allow for a direct computation in this limit. Hence, we refrain from giving an estimate for the scaling exponents and defer quantitative results to future publications.

 Let us close this investigation with a discussion of the spectral function in the critical regime. To begin with, for $k\to 0$ the pole contribution of the propagator vanishes as $Z_\phi \propto k^{-\eta_\phi} \to \infty$. In turn, for $k=0$, the scattering tail carries all the weight, and the solution for $k=0$ is given by 
\begin{align}\label{eq:Scalingrho}
\tilde\rho(\lambda)\propto \frac{1}{\lambda^{2\left( 1-\frac{\eta_\phi}{2}\right)}} \quad \to \quad G_k(p) \propto  \frac{1}{(p^2)^{\left( 1-\frac{\eta_\phi}{2}\right)}}  \,. 
\end{align}
Note that the scaling \labelcref{eq:Scalingrho} is naturally cut off in the infrared at $\lambda =2 m_\tinytext{pole}$ according to \labelcref{eq:scale-dep-spec}. In the ultraviolet, for $\lambda\to\infty$, the spectral function also has to decay faster than \labelcref{eq:Scalingrho}. As has been discussed in \cite{Bonanno:2021squ, Horak:2021pfr}, the propagators of \textit{physical} states or fields have to decay as $1/p^2$ for large momenta. This is at odds with \labelcref{eq:Scalingrho} and indeed the spectral function $\tilde \rho$ in \labelcref{eq:Scalingrho} is not (UV) normalisable. For a finite $k$, $\tilde\rho$ decays faster than $1/\lambda^2$, as is manifest in our explicit solutions in the broken and symmetric phase, \Cref{figPropSpecBroken,fig:PropSpecSym} respectively.

\section{Varying the truncation}\label{sec:classicVertices}

\begin{figure*}[t]
	\centering
	\begin{subfigure}{.49\linewidth}
	\centering
	\includegraphics[width=\textwidth]{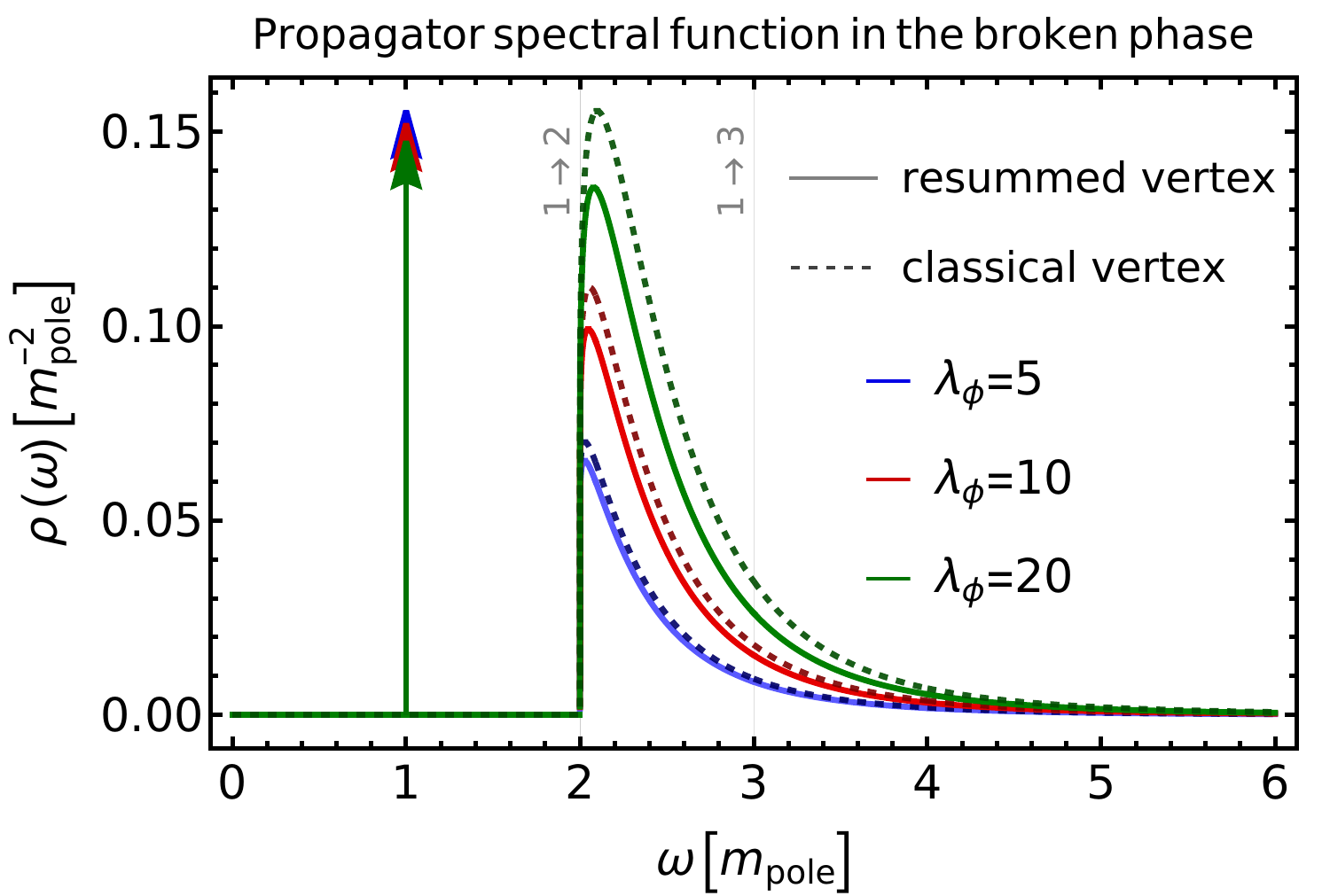}\vspace*{-0.5mm}
	\caption{Comparison of the spectral functions within a classical vertex approximation to the results with a bubble resummed vertex function. Both spectral functions where calculated via the spectral fRG. \hspace*{\fill}}
	\label{Resultpropagatorsclassic}
	\end{subfigure}\hspace*{0.05\linewidth}
	\begin{subfigure}{.47\linewidth}\vspace*{+0.5mm}
		\centering
		\includegraphics[width=\textwidth]{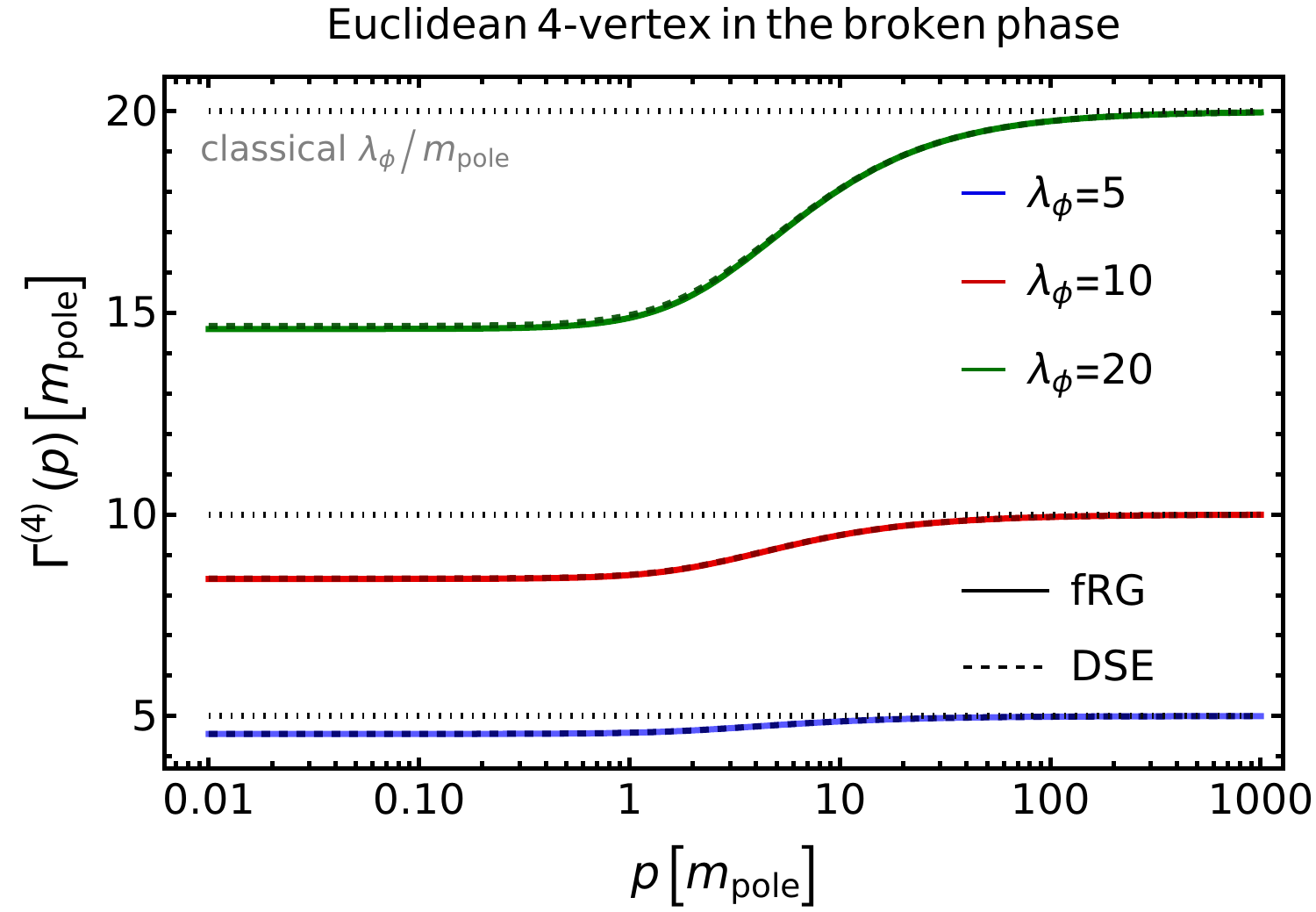}
		\subcaption{Four-point vertex as function of Euclidean frequency. Results are obtained from the respective spectral functions in \Cref{figPropSpecBroken}. The classical values of the vertices are indicated in grey.\hspace*{\fill}}
	\end{subfigure}
	\caption{Propagator spectral functions and euclidean four-point functions in the broken phase. All quantities are measured in units of the pole mass.\hspace*{\fill}}
\end{figure*} 
This section is dedicated to the comparison of our approximation to the classical vertex approximation. The latter is given by
\begin{align} \nonumber
\Gamma^{(4)}=&\lambda_{\phi} \,,\\[1ex] 
\Gamma^{(3)} =& \sqrt{3\lambda_{\phi}} m_\tinytext{curv}\, .
\end{align}
The value of the tadpole diagram is absorbed in the renormalisation constant and the only remaining contribution is the polarisation diagram~\labelcref{eq:polDiag}. Note that this approximation leads to a non-zero scattering spectrum only, if evaluated in the broken phase with a finite three-point function. 

The resulting spectral functions are presented in~\Cref{Resultpropagatorsclassic}, indicated by dashed lines. All curves are measured in units of the respective pole masses to compare the magnitude of the scattering spectra with our main results. By introducing the resummed quantum corrections to the four-point function, its amplitude in the infrared is lowered due to the negativity of the respective scattering spectra, see \Cref{figPropSpecBroken} and \Cref{fig:4PSpecbroken}. This leads to a systematic decrease of the spectral tail compared to the classical vertex approximation. Other than that, the visible structure is very similar. Nevertheless, the classical approximation misses any contribution of the tadpole and is quantitatively different to the result with a non-trivial four-point function for higher couplings in the broken phase. In the symmetric phase however, the inclusion of the tadpole momentum structure is crucial to generate a non-trivial scattering spectrum with the fRG.

\section{Technical details and numerics}\label{sec:Numerics}

This section is dedicated to the numerical solution of the flow equation. First, we rewrite the leading contributions as scale derivatives and integrate by parts in $k$-direction. This allows to flow the non-analyticities at the respective onset positions analytically, see \labelcref{eq:partialintegration}. We will make use of this relation to define consistent initial conditions in the UV. The second part explains the numerical algorithm we used to obtain the results given in \Cref{sec:Results}.

\subsection{Leading order and initial conditions}\label{sec:IC_leadingOrder}

Inserting \labelcref{eq:scale-dep-spec} in \labelcref{eq:generalDiags}, we find all combinations of poles and tails we have to integrate over. The leading order is given by the contribution of $\delta$-functions only and is already present on the classical level. To study the structure of the flow and the dependency of the result of the initial condition, we first note that certain contributions of the Callan-Symanzik flow can be rewritten in terms of a scale derivative. 
This is possible for every contribution to the Callan-Symanzik flow that carries only pole contributions on the two lines surrounding the (modified) regulator insertion in \Cref{Diag:flowingMinimum2Point}. This allows us to integrate the flow by parts, which reduces the degree of divergence of potential integrable singularities and simplifies the numerical treatment. To this end, we rewrite the (modified) fRG polarisation diagram at leading order as 
\begin{align}
 & I_\tinytext{pol}(m_\tinytext{pole},m_\tinytext{pole},m_\tinytext{pole},p^2)  =-\frac{1}{8k} \partial_k \tilde{I}_\tinytext{pol}(m_\tinytext{pole},m_\tinytext{pole},p^2)  \,,
\label{eq: firstorder factorisation}
\end{align}
where the factor $1/4k$ follows from the $k$-derivative of the spectral kernel with $\partial_k m_\tinytext{pole}^2=4k$ and another factor $1/2$ accounts for the double counting from hitting both arguments in k with the derivative. 
This connects both polarisation type momentum structures, $I_\tinytext{pol}$ and $\tilde{I}_\tinytext{pol}$, given in \labelcref{eq:Ipol}.
Evaluating every spectral parameter on the mass pole we can integrate the combined contribution of the polarisation and fish diagram to arrive at
\begin{align}\nonumber 
&\hspace{.5cm}\left[\Gamma^{(2)}(p^2)\right]^k_\Lambda\\[1ex]\nonumber
&=\int^k_{\Lambda}\frac{dk}{k} \; \frac{(\Gamma_{k}^{(3)})^2}{Z_\phi^2} \left(-\frac{\dot{\mathcal{R}}}{8Z_\phi k} \partial_k +\frac{A}{\phi_0} \right)\tilde{I}_\tinytext{pol}(p^2)\\[1ex]
&=\left[ - \mathcal{F}(k) \tilde{I}_\tinytext{pol}(p^2) \right]^{k}_{\Lambda} + \int^k_{\Lambda}\frac{dk}{k} \mathcal{L}(k)\tilde{I}_\tinytext{pol}(p^2)  \; , 
\label{eq:partialintegration}
\end{align}
where we summarised the prefactors of the fRG polarisation diagram and the prefactors from the $t$-integral as 
\begin{align}
\mathcal{F}(k)&= \frac{\dot{\mathcal{R}} \, (\Gamma_{k}^{(3)})^2 }{8\,Z_\phi^3 \, k^2} 
&\hspace{-0.5cm} = \left[\frac{(2-\eta_\phi)}{4 } + \frac{\mathcal{S}}{Z_\phi}\right] \frac{(\Gamma_{k}^{(3)})^2}{Z^2_\phi}\, ,
\end{align}
with
\begin{align}
\mathcal{S} = \frac{3}{16} \frac{m_\tinytext{curv}^2}{k^2}  \left(\frac{\partial_t \Delta m_\tinytext{curv}^2}{m_\tinytext{curv}^2} +\frac{\partial_t\Gamma^{(4)}}{\Gamma^{(4)}}\right) \, .
\end{align}
The boundary term will be the leading contribution. At one-loop order, \ie $Z_\phi=1$ and $\Gamma^{(4}=\lambda_\phi$, $\mathcal{S}$ vanishes identically and $\mathcal{F}$ reduces to $\frac12 (S^{(3)})^2$, which is the prefactor of the one-loop polarisation diagram times the squared classical three-vertex. The factor in the remaining integral reads 
\begin{align}\nonumber
 &\mathcal{L}(k)=\left(\partial_t\mathcal{F}(k) + \frac{A}{\phi_0}\frac{(\Gamma_{k}^{(3)})^2}{Z_\phi^2}\right) \\*[1ex]\nonumber
 & \, = \left\lbrace\frac{(2-\eta_\phi)}{4}\frac{\partial_t \Gamma^{(4)}}{\Gamma^{(4)}} - \frac{(2-\eta_\phi)}{2}\eta_\phi + \frac{\dot{\eta_\phi}}{4} - \frac{\eta_\phi\,\partial_t m_\tinytext{curv}^2}{4m_\tinytext{curv}^2}\right.  \\*[1ex] 
 &\, \left. +\frac{1}{Z_\phi}\left[ \partial_t  \mathcal{S} + \mathcal{S} \left(3\eta_\phi+\frac{\partial_t m_\tinytext{curv}^2}{m_\tinytext{curv}^2} +\frac{\partial_t\Gamma^{(4)}}{\Gamma^{(4)}}\right)\right]\right\rbrace \frac{(\Gamma_{k}^{(3)})^2}{Z_\phi^3}\,,
\end{align}
where the tree-level terms stemming from $\mathcal{F}(k)$ and $A$ cancel exactly. With that, we recover one-loop perturbation theory. Without the additional one-loop structure of the three-point function in \labelcref{eq:FullFlowGammaFinal1}, \ie $A=0$, the remaining tree-level term would spoil the one-loop result. 

To discuss the necessity of a consistent initial condition, it is instructive to work out the one-loop result from a spectral fRG perspective. In the large k limit, we can neglect the non-trivial flow of $\Gamma^{(2)}(p=0)$ and $Z_\phi$, leading to $Z_\phi=1$ and $(\Gamma_{k}^{(3)})^2=3\lambda m_\tinytext{pole}^2$ with $m_\tinytext{pole}^2= 2k^2$. \Cref{eq:partialintegration} is then readily integrated and reduces to 
\begin{align}\label{eq:oneloop}
 \Gamma_{1\text{-loop}}^{(2)}(p^2)\big|_k =&\,\Gamma^{(2)}(p^2)\big|_\Lambda +  3\lambda_\phi \left[k^2 \tilde{I}_\tinytext{pol}(\sqrt{2}k,\sqrt{2}k,p^2)\right]_k^{\Lambda}.
\end{align}
With a classical initial condition, \labelcref{eq:oneloop} leads to
\begin{align}
\label{eq:integratedOneloop}
&\Gamma_{1\text{-loop}}^{(2)}(p^2)\big|_k =p^2 +  \frac{3\lambda_\phi}{2} \left[2k^2 \tilde{I}_\tinytext{pol}(\sqrt{2}k,\sqrt{2}k,p^2)\right]_k^{\Lambda}\,.
\end{align}
Performing the Wick rotation of \labelcref{eq:integratedOneloop} and extracting the spectral function with \labelcref{eq:specrep}, we find that the one-loop scattering contribution to the spectral function is discontinuous at $\omega=2 \sqrt{2\Lambda^2}=2m_\tinytext{pole}^\Lambda$ and turns negative for larger spectral values. 
Clearly, leading order information is lost above the initial onset scale and can not be recovered by the flow. Even worse, for higher frequencies than $2 m_\tinytext{pole}^\Lambda$, the positivity of the spectral function is violated. This is cured by using RG-consistent initial conditions, which appears to be crucial to obtain a physical spectral function from the flow.
To that end, we require the solution to be independent of the initial scale $\Lambda$. This can be achieved, by sending the initial scale to infinity, corresponding to an initial condition that cancels the $\Lambda$ dependence trivially. This is done by choosing the initial condition to be an iterative solution of 
\begin{align}\nonumber
&\Gamma^{(2)}[ p^2,Z_{\phi},\Gamma^{(3)},m_\tinytext{pole}]= m_\tinytext{pole}^2 + p^2 - \frac{1}{2} \frac{(\Gamma^{(3)})^2}{Z_\phi^3}  \\[1ex] &\times \Bigl[ \tilde{I}_\tinytext{pol}(m_{\tinytext{pole}},m_{\tinytext{pole}},p^2) -\tilde{I}_\tinytext{pol}(m_{\tinytext{pole}},m_{\tinytext{pole}},-m_{\tinytext{pole}}^2)\Bigr] \, ,
\label{eq:initialcondition}
\end{align} 
where the last term accounts for the on-shell renormalisation. As initial guess we use the parameter of the classical effective potential \labelcref{eq:3vertex,eq:VeffUVbroken} with $Z_\phi=1$ and $m_\tinytext{pole}^2=2\Lambda^2$. In other words, we choose our initial condition to be compatible with \labelcref{eq:partialintegration}. Note that with this choice of initial conditions, the loss of leading order information is circumvented at all momentum scales, as all contributions of order $O(\lambda/k)$ are taken into account. The flow is initialised at large cutoff scales, where higher terms in $\lambda_\text{eff}=\lambda_\phi/k$ are strongly suppressed. To determine the three-point function dynamically,~\labelcref{eq:initialcondition} was coupled to the resummed four-point function via~\labelcref{eq:3vertex}. The initial values for $Z_{\phi}$ and $\Gamma^{(3)}$ are presented in \Cref{Tab:initialConditions}. 
\begin{table}\
\centering
\renewcommand{\arraystretch}{1.6}
\begin{tabular}{c|c|c}
$ \Lambda/\lambda_\phi$ & $\Gamma^{(3)}\;[\lambda_\phi^{\frac32}]$ &  $Z_{\phi}$ \\
\hline
$10 $ & $17.3084$ & $1.0007$ \\
\end{tabular}
\caption{Initial conditions obtained from \labelcref{eq:initialcondition}. We measure the initial RG-scale $\Lambda$ in units of the coupling and every other quantity in units of the mass. }
\label{Tab:initialConditions}
\end{table}

It is left to determine the flow of the vertices and $\eta$. These have exact diagrammatic expressions which are in parts necessary to consider. It is convenient to approximate $\partial_t \Gamma^{(4)}$ by the $t$-derivative of \labelcref{eq:resummed-four-pt-func}, where we only consider the contributions of the mass-pole for simplicity. It leads to 
\begin{align} \label{eq:gamma4dots}\nonumber
\partial_t \Gamma^{(4)}&\approx \frac{(1-2\eta_\phi)}{Z_\phi^2} \frac{ (\Gamma^{(4)})^2 }{16 \pi \, \sqrt{2}k}\,,\\[1ex]\nonumber
(\partial_t)^2 \, \Gamma^{(4)} &\approx \frac{ 2 (\partial_t \Gamma^{(4)})^2}{\Gamma^{(4)}}-\partial_t \Gamma^{(4)} \\[1ex] &+ \frac{(\eta_\phi(1-2\eta_\phi)-\dot\eta_\phi)}{Z_\phi^2} \frac{ (\Gamma^{(4)})^2 }{8 \pi \, \sqrt{2}k}\,.
\end{align}
The explicit k-dependences of \labelcref{eq:gamma4dots} can now be taken into account analytically in \labelcref{eq:partialintegration}. 
For $\eta_\phi$ we use the definition of $Z_\phi$ as the residue on the mass-pole
\begin{subequations}
\begin{align}
Z_\phi &=- \partial_{\omega^2} \Gamma^2(\omega^2)\big|_{\omega^2=m_\tinytext{pole}^2} \,.
\label{eq:Zphi}
\end{align}
With the parametrisation of the real part of the inverse propagator as
\begin{align}\label{eq:Gamma2parametrisation}
\Gamma^{(2)}(\omega^2)&=Z(\omega)\left( m_\tinytext{pole}^2-\omega^2 \right) \, ,
\end{align}
 the anomalous dimension $\eta_\phi$ is computed conveniently from the momentum derivative of the flow on the mass-pole,
\begin{align}\label{eq:eta}
\eta_\phi &=   \frac{1}{Z_\phi} \partial_{\omega^2} \frac{d}{dt}\Gamma^2(\omega^2)\big|_{m_\tinytext{pole}^2}  - \frac{1}{Z_\phi}4k^2 \partial_{\omega^2} Z(\omega) \big|_{m_\tinytext{pole}^2}\, .
\end{align}
\end{subequations}
Only the diagrams of~\labelcref{eq:final2pointflowREN} contribute due to the momentum derivative. The second term is given in terms of the spectral function:
\begin{align}
\partial_{\omega^2} Z(\omega) \big|_{\omega^2=m_\tinytext{pole}^2}= \frac{1}{Z_\phi}\int_\lambda \frac{\rho(\lambda)}{\lambda^2-m_\tinytext{pole}^2} \,.
\end{align}
The other parameters such as $\dot\eta_\phi$ and $\partial_t\Delta m_\tinytext{curv}^2$ where approximated by a numerical right-derivative.

\subsection{Numerical implementation} 

The numerical implementation uses Mathematica \cite{Mathematica}. The leading order contribution to the flow was integrated by means of \labelcref{eq:partialintegration}, where we split the explicit $k$-dependencies of each term from the sub-leading running of the respective parameter. This was facilitated by the split of the tree-level curvature mass: $m_\tinytext{curv}^2 = Z_\phi (2k^2) + \Delta m_\tinytext{curv}^2 $, as it allowed us to incorporate the tree-level running of the integrand in \labelcref{eq:partialintegration} analytically. The sub-leading corrections to the flow-parameter were approximated as constants in each step, while the combined $k$-dependence of $\tilde{I}_\tinytext{pol}$ and the tree-level $k$-dependence of $m_\tinytext{curv}^2$ and $\partial_t \Gamma^{(4)}$ was integrated analytically.

For the one-cut contributions, including the tadpole, we approximated the $k$-integral by an explicit Euler scheme. For the sake of computation time, higher dimensional spectral integrals where dropped as they were numerically negligible in the considered coupling range in comparison to the leading order and next to leading order contributions. For an investigation of the scaling limit, their incorporation is crucial. The numerical integrations of spectral integrals were carried out using a global adaptive integration strategy. All contributions to diagrams were calculated and interpolated separately, where we used finer grids around sharp structures and more coarse grids where the functions are smooth. 
We implemented a local feedback of the spectral function with a step size $dk=0.005$, using the spectral function $\rho$ to calculate $\partial_t \Gamma^{(2)}$. The correct renormalisation was enforced conveniently in every step by subtracting the value of the inverse propagator on the mass pole. The residue on the pole was extracted from $\Gamma^{(2)}(p)$ in each step via \labelcref{eq:Zphi}.

\section{Calculation of diagrams}\label{sec:analyticintegrands}

In this section, all diagrammatic expressions appearing in the main text are given in analytic form. The spectral approach we use, allows us to calculate diagrams with full propagators in terms of integrals known from perturbation theory. The insertion of a mass-derivative in \Cref{fig:2pointflow} leads to a squared propagator on one line in comparison with the usual vacuum polarisation or tadpole diagram. Using the spectral representation, the momentum structure of the regulator line can be rewritten via a partial fraction decomposition
\begin{align}\label{eq:partialfraction}
\frac{1}{(\lambda_1^2 + q^2)(\lambda_2^2 + q^2)}= \frac{-1}{(\lambda_1^2 - \lambda_2^2)} \left(\frac{1}{\lambda_1^2 + q^2}-\frac{1}{\lambda_2^2 + q^2}\right) \,.
\end{align} 
A given (spectral) flow-diagram can therefore be reduced to the computation of the momentum integral, where the regulator line is replaced by a single propagator, which we will denote with a tilde. Denoting the spectral parameter of the divided line as $\lambda_1$ and $\lambda_2$, we write schematically
\begin{align}\nonumber 
\text{D}(\lambda_1,\lambda_2, ...,p^2)&  \\[1ex]
& \hspace{-2cm}= \frac{-1}{(\lambda_1^2 - \lambda_2^2)}\,\Bigl[\tilde{D}(\lambda_1 ...,p^2)-\tilde{D}(\lambda_2 ...,p^2)\Bigr] \, ,
\end{align}
and accordingly
\begin{align}\label{equ:symmetricIntegrands}
\text{D}(\lambda,\lambda, ...,p^2)= \frac{-1}{2\lambda}\partial_{\lambda} \, \tilde{D}(\lambda ...,p^2) \, .
\end{align}
This reduces the calculation of $I_\text{pol}$, as defined in \labelcref{eq:Ipol} to the calculation of $\tilde{I}_\tinytext{pol}$ as given below
\begin{align}\nonumber 
&I_\tinytext{pol}(\lambda_1,\lambda_2,\lambda_3,p^2)\\[1ex]
&=  \frac{-1}{(\lambda_1^2 - \lambda_2^2)} \Bigl[\tilde{I}_\tinytext{pol}(\lambda_1,\lambda_3,p^2)-
\tilde{I}_\tinytext{pol}(\lambda_2,\lambda_3,p^2) \Bigr]\, ,
\label{eq:polDiag}\end{align}
Denoting the Euclidean and Minkowskian frequencies by $p$ and $\omega$ respectively, the momentum structure of the spectral polarisation diagram reads
\begin{align}\nonumber 
\tilde{I}_\tinytext{pol}(p,\lambda_1,\lambda_2) =& \frac{1}{4 \pi p} \textrm{Arctan}\left[\frac{p}{\lambda_1 + \lambda_2}\right]\,, \\[1ex]\nonumber
\tilde{I}_\tinytext{pol}(\omega,\lambda_1,\lambda_2) =& \frac{1}{4 \pi \omega} \Biggl\{\textrm{Arctanh}\left[\frac{w}{\lambda_1 + \lambda_2}\right] \\[1ex]
 &\hspace{1cm}+ \imag \, \theta\big(\omega - (\lambda_1 + \lambda_2)\big)\Biggr\} \,, 
\label{eq:polDiagDSEmink}
\end{align}
%
\begin{figure}[h!]
	\centering
	\includegraphics[width=0.46\textwidth]{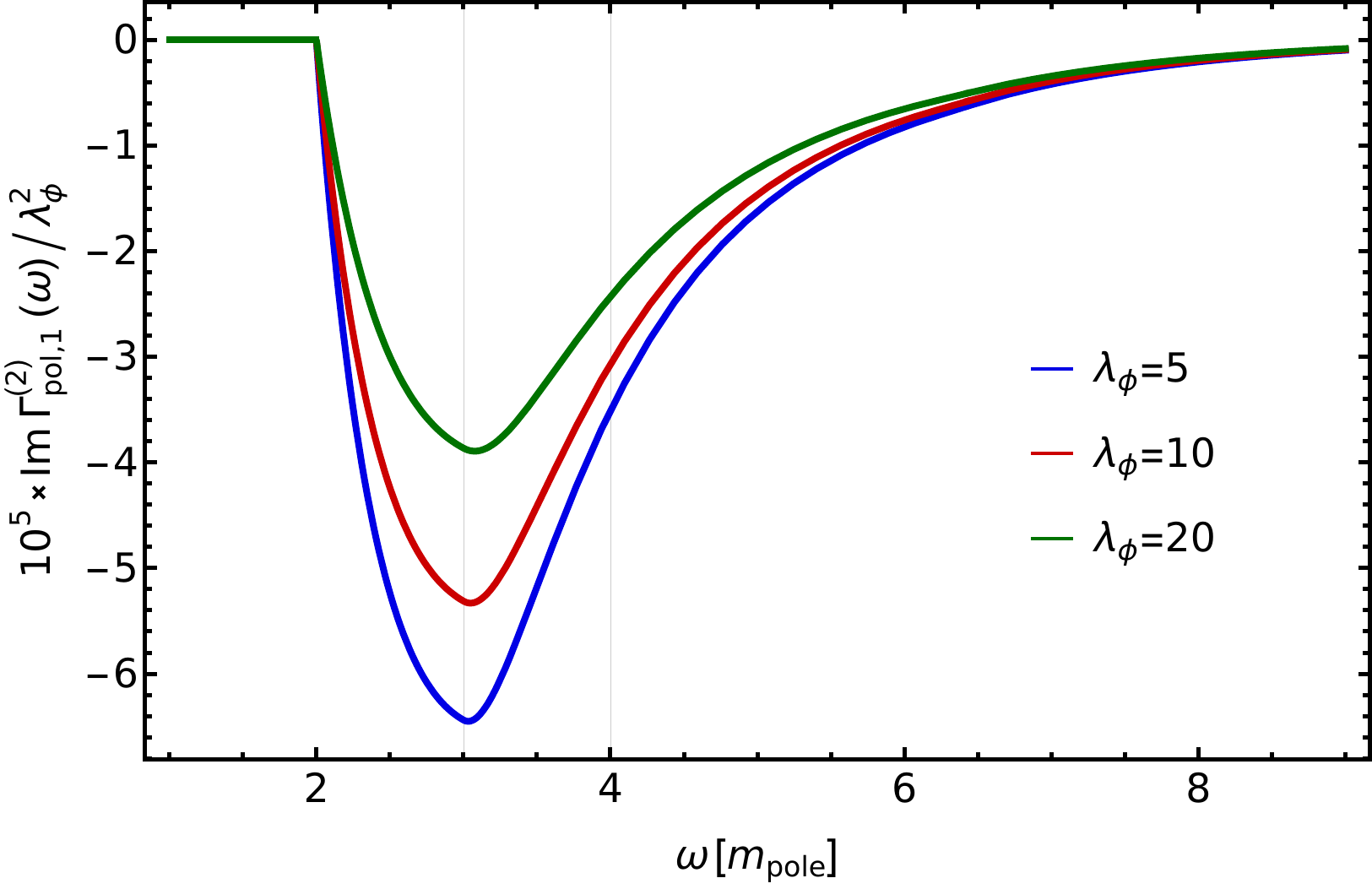}\vspace{0.5cm}
	\hspace*{-3mm}\includegraphics[width=0.479\textwidth]{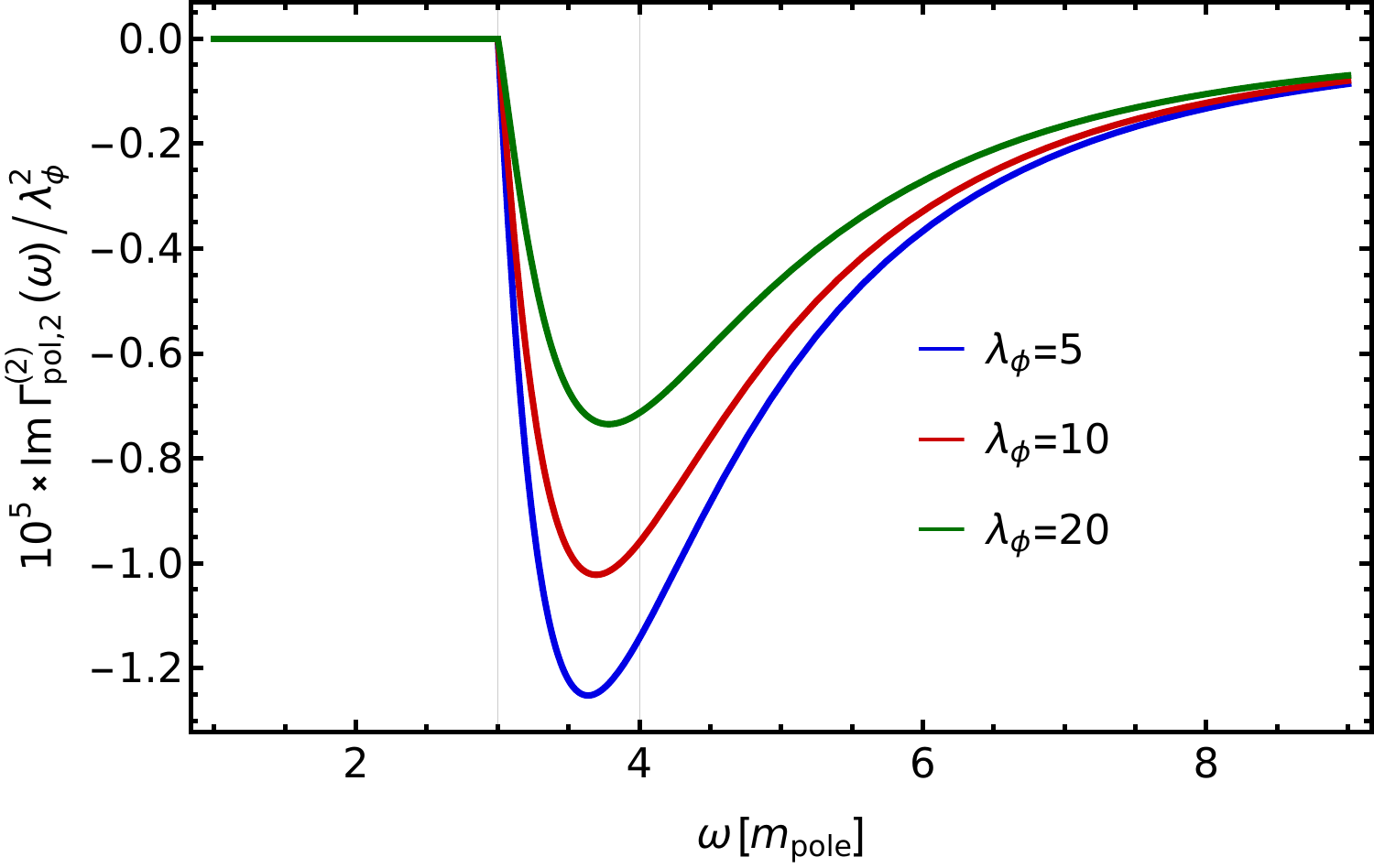}\vspace{0.5cm}
	\includegraphics[width=0.46\textwidth]{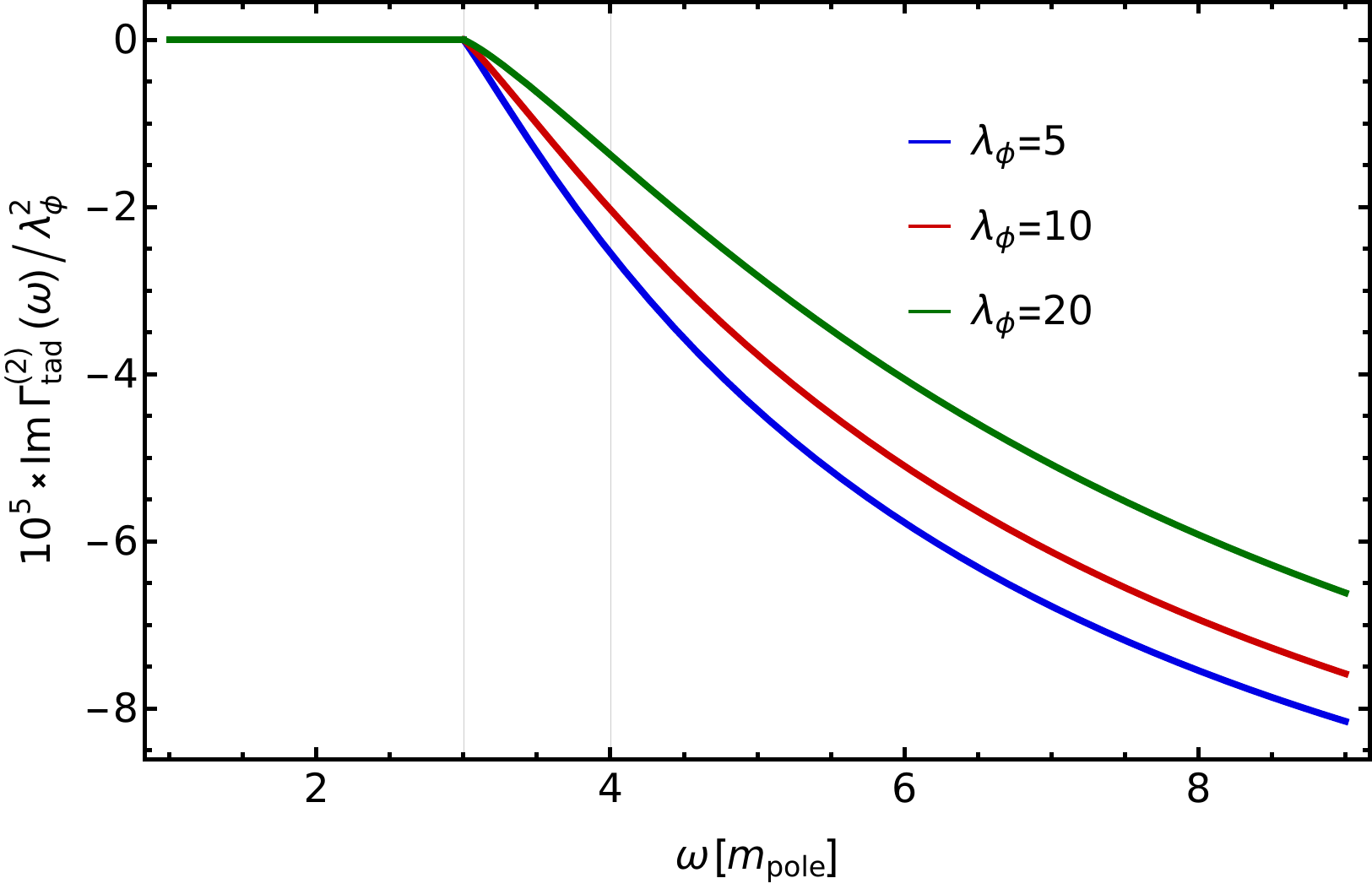}
	\caption{Next to leading order contributions to the imaginary part of the integrated flow in the broken phase rescaled by $\lambda_\phi^2$. All dimension-full quantities are measured in units of the pole mass. The first two figures show the contributions of polarisation diagram. They start at $2 m_\tinytext{pole}$ and $3 m_\tinytext{pole}$ respectively. The third figure is the dominating tadpole contribution, \ie, the insertion of $\rho_4$ for the vertex and only the $pole$ contributions for the propagators. It exhibits only a three-particle onset as discussed in \Cref{sec:Resultssymmetric}. \hspace*{\fill}}
	\label{fig:highercorrectionsresummed}
\end{figure}
%
see \cite{Horak:2020eng, Rajantie:1996np}. We find the integrand $I_\tinytext{pol}$ of the polarisation diagram to have a branch cut with compact support, i.e., for $\omega \in \left[\lambda_1 + \lambda_3, \lambda_2 + \lambda_3\right]$ for $\lambda_1 \leq \lambda_2$. This peculiarity is a dimension dependent property of the polarisation diagram which does not hold in four dimensions. There, the imaginary part of $I_{\text{pol}}$ has support for $\omega \in [\lambda_1 + \lambda_3, \infty)$ for $\lambda_1 \leq \lambda_2$. 

The onset position of these structures allow us to discuss the scattering continua. To this end we note that the diagrams with a polarisation topology have two or three lines that can carry either a mass-pole or a scattering contribution, see \Cref{Diag:flowingMinimum2Point}. If all lines carry a pole contribution, which is the leading order for all couplings in the considered coupling range, we find the flow of a discontinuity seeded at $2m_\tinytext{pole}$, representing a $1\rightarrow 2$ scattering.

\section{Higher order contributions to $\text{Im} \, \Gamma^{(2)}$ }
\label{sec:higherorderplots}

In this section we discuss the feedback of the scattering continuum, and present the next to leading order contributions. We restrict ourselves to the imaginary part, since it carries the dominant features. In contrast, the contribution of the respective real parts to the spectral function cannot be separated from each other as they only appear in the denominator of \labelcref{eq:rho}. 
Inserting exactly one scattering continuum in one of the top lines of the polarisation diagram, we find a contribution adding to both the $1\rightarrow2$ and the $1\rightarrow 3$ onset, consequently also starting at $2 m_\tinytext{pole}$ as can be seen in the top panel of~\Cref{fig:highercorrectionsresummed}. The dominant sharp onsets of the $1\rightarrow 3$ scattering are found in the polarisation diagram by inserting exactly one scattering continuum into the bottom line and into the tadpole. Their integrated flows are presented in the middle and lower panel of \Cref{fig:highercorrectionsresummed}. Note, that they correspond to different diagrammatic topologies. The $1\rightarrow 3$ contribution of the polarisation diagram can be described diagrammatically by two consecutive $1 \rightarrow 2$ scatterings whereas the tadpole reproduces a sunset topology.
The insertion of two or three scattering continua lead to sharp, although strongly suppressed onsets at $3m_\tinytext{pole}$ and $4m_\tinytext{pole}$.

In~\Cref{fig:highercorrectionsresummed}, the one-cut corrections to the imaginary part are given in units of their respective coupling strength. The re-scaled contributions are qualitatively compatible with each other, showing the proportionality of the one-cut-contribution to $\lambda_{\phi}^2$. This can be anticipated from two-loop perturbation theory. The decrease of the peak for higher couplings is connected to the decrease of the residue $Z^{-1}$ of the mass pole, as shown in \Cref{figPropSpecBroken}. 
The tadpole contribution is at leading order proportional to $\lambda_{\phi}^2$. On a perturbative level, the first dynamic contribution is introduced by the first bubble diagram in \Cref{fig:bubble}. This is confirmed in the third panel of \Cref{fig:highercorrectionsresummed}, where the different tadpole contributions share the same order of magnitude if rescaled with $\lambda_{\phi}^2$.

\section{Flow equation of the effective potential} 
\label{sec:CS-EffPot}

In this section we briefly discuss the flow equation of the effective potential in the local potential approximation for the sake of completeness and for the illustration of consistency of the approach. Its derivation including the determination of the counter term has been discussed in detail in Appendix~A of \cite{Braun:2022mgx}. The flow of the first field derivative of the effective potential in three dimensions is given by 
\begin{align}\nonumber 
	\partial_\mu 	V_\textrm{eff}^{(1)}(\phi) =&\, -\frac12 \int \frac{\textrm{d}^3 p}{(2 \pi)^3} \, 
	\frac{  V^{(3)}_\textrm{eff}(\phi) }{ \left[ p^2 + 	V^{(2)}_\textrm{eff}(\phi) \right]^2 }  \\[1ex]  
	&\,+\phi - \partial_\mu   S^{(1)}_{\text{ct}}[\phi]  \,, 
\label{eq:muFlowV'}
\end{align}
where we have dropped the multiplication with $\mu$ present in \labelcref{eq:CSflow-gamma}. 
We have already used that the momentum integral in \labelcref{eq:muFlowV'} is finite, and hence we can remove additional regularisations such as dimensional regularisation relevant in the $d=4$ case, see again Appendix~A of \cite{Braun:2022mgx}. The momentum integral in \labelcref{eq:muFlowV'} is readily performed and we arrive at 
\begin{align}
	\partial_\mu 	V^{(1)}_\textrm{eff}(\phi) = -\frac{1}{4\pi^2} \frac{ V^{(3)}_\textrm{eff}(\phi) }{\sqrt{ V^{(2)}_\textrm{eff}(\phi) }} +\phi  -  \partial_\mu   S^{(1)}_{\text{ct}}[\phi]  \,, 
	\label{eq:muFlowV'analytic}
\end{align}
and upon $\phi$-integration we are led to  
\begin{align}
	\partial_\mu 	V_\textrm{eff}(\phi) = -\frac{1}{8\pi^2}  \sqrt{ V^{(2)}_\textrm{eff}(\phi) } +\frac12 \phi^2 -  \partial_\mu   S_{\text{ct}}[\phi] \,,  
	\label{eq:muFlowV}
\end{align}
where we have set the integration constant to zero. Note that \labelcref{eq:muFlowV} has a peculiar form: the loop contribution is negative, while its diagrammatic form is seemingly positive but not well-defined without regularisation. We emphasise that the the first field derivative of the flow \labelcref{eq:muFlowV} is negative (times $V^{(3)}_\textrm{eff}$), see \labelcref{eq:muFlowV'analytic}, as holds true for all momentum-cutoff flows.

It is illustrative to consider the large field limit with $\phi^2/|\mu| \to \infty$. For these field values the effective potential 
(or rather its interaction part) reduces to the classical one, and the flow reduces to 
\begin{align}
\partial_\mu 	V_\textrm{eff}(\phi) \to  -\frac{1}{8\pi^2}  \sqrt{\frac{\lambda_\phi}{2} \phi^2 } +\frac12 \phi^2 -  \partial_\mu   S_{\text{ct}}[\phi]  \,, 
	\label{eq:muFlowVLargephi}
\end{align}
up to sub-leading terms. We note in passing that \labelcref{eq:muFlowVLargephi} shows the self-consistency of the assumption that the interaction part reduces to the classical one. The right hand side is proportional to $|\phi| =\sqrt{2 \rho}$. This reflects the infrared cut in three dimensional momentum cutoff flows for $\mu\to 0$. For the CS flow it is present for all $\mu$ in the large field limit in contradistinction to momentum cutoff flows that decay with $1/ V_\textrm{eff}^{(2)}(\phi)$ for large fields.

\end{document}